\documentclass[journal ]{new-aiaa}
\usepackage[utf8]{inputenc}
\usepackage{textcomp}

\usepackage{graphicx}
\usepackage{amsmath}
\usepackage{float}
\usepackage{subcaption}
\usepackage[version=4]{mhchem}
\usepackage{siunitx}
\usepackage{longtable,tabularx}
\usepackage[table]{xcolor}
\usepackage{overpic}
\usepackage{multirow}

\definecolor{a_color}{RGB}{68, 1, 84}
\definecolor{b_color}{RGB}{72, 35, 116}
\definecolor{c_color}{RGB}{60, 68, 151}
\definecolor{d_color}{RGB}{48, 102, 141}
\definecolor{e_color}{RGB}{42, 129, 135}
\definecolor{f_color}{RGB}{41, 149, 121}
\definecolor{g_color}{RGB}{72, 167, 87}
\definecolor{h_color}{RGB}{118, 183, 46}
\definecolor{i_color}{RGB}{174, 192, 16}
\definecolor{j_color}{RGB}{225, 185, 37}
\definecolor{k_color}{RGB}{254, 152, 41}
\definecolor{l_color}{RGB}{253, 186, 34}
\definecolor{m_color}{RGB}{254, 232, 31}
\definecolor{n_color}{RGB}{254, 255, 0}
\definecolor{o_color}{RGB}{202, 244, 47}
\definecolor{p_color}{RGB}{250, 127, 20}
\definecolor{q_color}{RGB}{80, 191, 230}
\definecolor{r_color}{RGB}{140, 15, 155}

\definecolor{a_color_3d}{RGB}{68, 1, 84}
\definecolor{b_color_3d}{RGB}{72, 33, 117}
\definecolor{c_color_3d}{RGB}{54, 86, 136}
\definecolor{d_color_3d}{RGB}{41, 120, 142}
\definecolor{e_color_3d}{RGB}{37, 149, 135}
\definecolor{f_color_3d}{RGB}{77, 157, 123}
\definecolor{g_color_3d}{RGB}{106, 176, 114}
\definecolor{h_color_3d}{RGB}{138, 189, 101}
\definecolor{i_color_3d}{RGB}{175, 200, 0} 
\definecolor{j_color_3d}{RGB}{210, 160, 0} 
\definecolor{k_color_3d}{RGB}{240, 130, 90} 
\definecolor{l_color_3d}{RGB}{253, 151, 92}
\definecolor{m_color_3d}{RGB}{253, 131, 97}
\definecolor{n_color_3d}{RGB}{252, 104, 110}
\definecolor{o_color_3d}{RGB}{249, 71, 122}
\definecolor{p_color_3d}{RGB}{236, 52, 138}
\definecolor{q_color_3d}{RGB}{217, 32, 146}
\definecolor{r_color_3d}{RGB}{188, 19, 153}
\definecolor{s_color_3d}{RGB}{154, 10, 160}
\definecolor{t_color_3d}{RGB}{117, 1, 128}
\definecolor{u_color_3d}{RGB}{40, 60, 150}
\definecolor{v_color_3d}{RGB}{0, 150, 150}

\newcommand{\aTwoD}{ \cellcolor{a_color!40}a }
\newcommand{\bTwoD}{ \cellcolor{b_color!40}b }
\newcommand{\cTwoD}{ \cellcolor{c_color!40}c }
\newcommand{\dTwoD}{ \cellcolor{d_color!40}d }
\newcommand{\eTwoD}{ \cellcolor{e_color!40}e }
\newcommand{\fTwoD}{ \cellcolor{f_color!40}f }
\newcommand{\gTwoD}{ \cellcolor{g_color!40}g }
\newcommand{\hTwoD}{ \cellcolor{h_color!40}h }
\newcommand{\iTwoD}{ \cellcolor{i_color!40}i }
\newcommand{\jTwoD}{ \cellcolor{j_color!40}j }
\newcommand{\kTwoD}{ \cellcolor{k_color!40}k }
\newcommand{\lTwoD}{ \cellcolor{l_color!40}l }
\newcommand{\mTwoD}{ \cellcolor{m_color!40}m }
\newcommand{\nTwoD}{ \cellcolor{n_color!40}n }
\newcommand{\oTwoD}{ \cellcolor{o_color!40}o }
\newcommand{\pTwoD}{ \cellcolor{p_color!40}p }
\newcommand{\qTwoD}{ \cellcolor{q_color!40}q }
\newcommand{\rTwoD}{ \cellcolor{r_color!40}r }

\newcommand{\smallcellcolor}[2]{\colorbox{#1}{\makebox(3,3){$\scriptstyle #2$}} }

\newcommand{\aThreeD}{ \smallcellcolor{a_color_3d!40}{a} }
\newcommand{\bThreeD}{ \smallcellcolor{b_color_3d!40}{b} }
\newcommand{\cThreeD}{ \smallcellcolor{c_color_3d!40}{c} }
\newcommand{\dThreeD}{ \smallcellcolor{d_color_3d!40}{d} }
\newcommand{\eThreeD}{ \smallcellcolor{e_color_3d!40}{e} }
\newcommand{\fThreeD}{ \smallcellcolor{f_color_3d!40}{f} }
\newcommand{\gThreeD}{ \smallcellcolor{g_color_3d!40}{g} }
\newcommand{\hThreeD}{ \smallcellcolor{h_color_3d!40}{h} }
\newcommand{\iThreeD}{ \smallcellcolor{i_color_3d!40}{i} }
\newcommand{\jThreeD}{ \smallcellcolor{j_color_3d!40}{j} }
\newcommand{\kThreeD}{ \smallcellcolor{k_color_3d!40}{k} }
\newcommand{\lThreeD}{ \smallcellcolor{l_color_3d!40}{l} }
\newcommand{\mThreeD}{ \smallcellcolor{m_color_3d!40}{m} }
\newcommand{\nThreeD}{ \smallcellcolor{n_color_3d!40}{n} }
\newcommand{\oThreeD}{ \smallcellcolor{o_color_3d!40}{o} }
\newcommand{\pThreeD}{ \smallcellcolor{p_color_3d!40}{p} }
\newcommand{\qThreeD}{ \smallcellcolor{q_color_3d!40}{q} }
\newcommand{\rThreeD}{ \smallcellcolor{r_color_3d!40}{r} }
\newcommand{\sThreeD}{ \smallcellcolor{s_color_3d!40}{s} }
\newcommand{\tThreeD}{ \smallcellcolor{t_color_3d!40}{t} }
\newcommand{\uThreeD}{ \smallcellcolor{u_color_3d!40}{u} }
\newcommand{\vThreeD}{ \smallcellcolor{v_color_3d!40}{v} }

\usepackage{tikz}

\usetikzlibrary{arrows.meta}
\usetikzlibrary{calc}
\usetikzlibrary{external}
\usetikzlibrary{positioning}

\renewcommand{\arraystretch}{0.65}
\newcommand{\genOfG}{\{\text{generators of $G$}\}}
\newcommand{\generate}[1]{\langle #1\rangle}
\newcommand{\R}{\mathbb{R}}
\newcommand{\x}{\mathrm{x}}
\newcommand{\Layer}{\mathcal{L}}
\newcommand{\Affin}{\mathcal{A}}
\newcommand{\Perm}{\mathbf{P}}
\newcommand{\f}{\mathbf{f}}

\newenvironment{xsmallmatrix}
  {\renewcommand\thickspace{\kern0em}
   \smallmatrix}
  {\endsmallmatrix}

\title{Enhancing lattice kinetic schemes for fluid dynamics with Lattice-Equivariant Neural Networks}

\author{Giulio Ortali 
        \footnote{Ph.D. Candidate, Eindhoven University of Technology and SISSA Trieste, g.ortali@tue.nl} 
       }
       \affil{Eindhoven University of Technology, 5600 MB Eindhoven,~ The Netherlands}
       \affil{SISSA (International School for Advanced Studies), Trieste,~ Italy}

\author{Alessandro Gabbana
        \footnote{Postdoctoral Researcher, Los Alamos National Laboratory, agabbana@lanl.gov}
        }
        \affil{CCS-2 Computational Physics and Methods, Los Alamos National Laboratory, Los Alamos, 87545 New Mexico, USA}
        \affil{Center for Nonlinear Studies (CNLS), Los Alamos National Laboratory, Los Alamos, 87545 New Mexico, USA}

\author{Imre Atmodimedjo
       \footnote{Master Student, Eindhoven University of Technology, i.s.atmodimedjo@student.tue.nl}
       }\affil{Eindhoven University of Technology, 5600 MB Eindhoven,~ The Netherlands}
\author{Alessandro Corbetta
        \footnote{Assistant Professor, Eindhoven University of Technology, a.corbetta@tue.nl}
        }
        \affil{Eindhoven University of Technology, 5600 MB Eindhoven,~ The Netherlands}
        \affil{Eindhoven Artificial Intelligence Systems Institute, 5600 MB Eindhoven,~ The Netherlands}

\begin{document}

\maketitle

\begin{abstract}

We present a new class of equivariant neural networks, hereby dubbed Lattice-Equivariant Neural Networks (LENNs), designed to satisfy local symmetries 
of a lattice structure. Our approach develops within a recently introduced framework aimed at learning neural network-based surrogate models  Lattice Boltzmann collision operators. Whenever neural networks are employed to model physical systems, respecting symmetries and equivariance properties has been shown to be key for accuracy, numerical stability, and performance.

Here, hinging on ideas from group representation theory, we define trainable layers whose algebraic structure is equivariant with respect to the symmetries of the lattice cell. Our method naturally allows for efficient implementations, both in terms of memory usage and computational costs, supporting scalable training/testing for lattices in two spatial dimensions and higher, as the size of symmetry group grows. We validate and test our approach considering 2D and 3D flowing dynamics, both in laminar and turbulent regimes. We compare with group averaged-based symmetric networks and with plain, non-symmetric, networks, showing how our approach unlocks the (a-posteriori) accuracy and training stability of the former models, and the train/inference speed of the latter networks (LENNs are about one order of magnitude faster than group-averaged networks in 3D). Our work opens towards practical utilization of machine learning-augmented Lattice Boltzmann CFD in real-world simulations.

\end{abstract}

\newpage

\section{Introduction}\label{sec:intro}
Machine learning (ML) algorithms have emerged as powerful tools for modeling complex phenomena, 
finding application in different engineering and science areas: from accelerating numerical methods, to offering unprecedented opportunities for extracting physical laws from ever-growing availability of extensive datasets, experimental and numerical~\cite{osti_1852843,RevModPhys.91.045002}.
Although purely data-driven models may be very successful in fitting observational data, their predictions generally
lack physical consistency, severely hampering generalization capabilities, possibly reliability, and their widespread adoption 
in real-world applications.
Designing ML algorithms complying with physical laws, which can often be phrased in terms of symmetry properties, as invariance and equivariance constraints,  is an active field of research and the main goal of the so-called Physics-Informed Machine Learning (PIML)~\cite{karniadakis-natrev-2021}.
More in general, symmetries are deeply interconnected with conservation laws~\cite{noether-1918}, and it is therefore highly desirable for any machine learning model applied to a physical system to adhere to such properties. 

Formally, the requirement that a predictive model, $f$, is invariant with respect to all elements of a given  transformation group $g\in G$ can be expressed as $f(\x) = f ( g (\x) ), \ \forall \x,\ \forall g \in G$; here $\x$ denotes the model input. Elements $g\in G$ could be, for instance, shifts in time (energy conservation) or space roto-translation (Galilean invariance). Indistinguishability of particles in a set of $N$ elements can also be cast in this framework, as invariance to permutations. In this work we are concerned with the more general notion of equivariance. Heuristically, this means that applying a transformation $g$ prior or after the evaluation of $f$ on a datum $\x$ must yield the same result; i.e., in formulas:  
\begin{align}\label{eq:eq}
  g(f(\x)) = f ( g (\x) ) , \quad \forall \x,\ \forall g \in G.
\end{align}
The relation in Eq.~\ref{eq:eq} requires that input and output spaces of $f$ coincide (this constraint can be further relaxed, cf. Ref.~\cite{cohen-icml-2016}).
Various approaches have been proposed for incorporating properties such as Eq.~\ref{eq:eq} into ML models,
with geometric deep learning  
being the framework specialized in addressing symmetries in the Artificial Neural Networks (ANN) context (see Ref.~\cite{bronstein-arxiv-2021} for a thorough introduction on the topic).
A simple approach for incorporating symmetries in ML models is data augmentation, which involves enlarging the annotated dataset used for model training with modifications of the input data through the actions of the symmetry group $G$~\cite{cubuk-arxiv-2018, lorraine-icai-2020}. 
Another possibility, known as group averaging~\cite{laptev-ieee-2016}, entails averaging the network predictions over transformations of the input via all the possible elements of the transformation group, in formulas  $\sum_{g \in G } g^{-1} \mathcal{F} (g \x)$ for a network $\mathcal{F}$ and an input $\x$.
The former is arguably the most straightforward approach, but leaves the ANN with the task of learning symmetries from data, resulting in reduced training efficiency and the impossibility to guarantee equivariance of the model during testing - key, e.g., in numerical simulation. On the opposite, the latter injects an inductive bias in the network architecture, ensuring that all the 
invariant transformations taken into consideration are satisfied by construction. Both methods scale poorly 
in higher dimensions and in general for large transformation groups.
Networks have been also internally engineered for symmetry. The most notable examples are 
convolutional neural networks (CNNs)~\cite{fukushima-bc-1980,lecun-nc-1989,NIPS2012_c399862d}
that embed translational symmetry, and have revolutionized computer vision.
This approach has been generalized to include other symmetries in group equivariant 
CNNs~\cite{cohen-icml-2016, ravanbakhsh-icml-2017, kondor-icml-2018, weiler-arxiv-2021}, 
which take into account not just translational, but also e.g. rotational and reflection symmetry. 
Further generalizations~\cite{finzi-icml-2021} have expanded group equivariant CNNs toward arbitrary symmetry groups. 
While CNNs require structured grid data, a more general and flexible approach is offered 
by graph neural networks (GNNs)~\cite{wu-ieee-2020}, which allow to 
embed invariance properties when dealing with unstructured discretization, e.g. in terms of particles~\cite{keriven-anips-2019}.

The current work develops in the context of accelerating lattice-based numerical simulations of fluids, using ANN models as surrogates. 
Within lattice-based fluid dynamic solvers, the Lattice Boltzmann Method~\cite{succi-book-2018}, 
hinged on discrete kinetic equations formulated at the mesoscopic scale, is a key representative in view of its computational efficiency,
and support for the simulation of a wide range of complex flows.
To date, few works have discussed the possibility of enhancing LBM by means of ANN,
with most of the efforts focused on accelerating the calculation of steady-state flows
using CNN~\cite{hennigh-arxiv-2017, guo-proc-2016, wang-tpm-2021}.
More recently, there have been more general attempts of employing ANNs for learning novel 
collision operators~\cite{corbetta-epje-2023}, tuning 
the parameters of already established models~\cite{bedrunka-hpc-2021,horstmann-cf-2024},
and even learning new subgrid scale modeling of turbulent flows~\cite{ortali-arxiv-2024}.
Specifically, the framework introduced in~\cite{corbetta-epje-2023} has established the issue of accurately learning the traditional BGK collisions as a quality benchmark for methods aiming at learning generic operators. That study has shown that respecting the local symmetries of the lattice is key for accuracy and numerical stability. There, symmetry has been achieved by means of networks group-averaged over the symmetry group of the lattice, with experiments limited to the 2D context. Group average-based strategies  exhibit poor scalability with respect to the size of the symmetry group up to the point that they are infeasible in 3D, where the symmetry group of the lattice grows by almost an order of magnitude.

In this work, we overcome the scalability limit of group-averaged approaches and propose a new class of neural networks, tailored for lattice-based computations and exposing built-in local symmetries.  We dub them Lattice-Equivariant Neural Networks (LENN). The trainable weights 
of each single layer of our architecture are constrained to comply exactly with equivariant properties connected with the lattice symmetry, and unlock ANN-based equivariant surrogates that can effectively operate also in three dimensional lattices (and possibly higher dimension).

We compare accuracy and computational efficiency of training and validation of LENNs against 
a plain (non-equivariant) Multilayer Perceptron (MLP) architecture, and a MLP trained with the group averaging approach.
We consider both laminar (2D, 3D) and turbulent settings (3D), confirming that retaining 
equivariant properties is key for systematically reproducing the time dynamics of fluid flows.
We show that our method achieves results comparable to those of group averaging in terms of accuracy, yet at a significant lower computational cost.

This work is organized as follows.
In Sec.~\ref{sec:methods} we provide an overview of the numerical methods employed in this work, starting from a 
brief overview on ANN, followed by a description of LBM and the definition of the problem of learning a collision operator. Group-averaging-based methods are also reviewed. 
In Sec.~\ref{sec:lenn} we introduce our novel LENN architecture, which we test and evaluate in Sec.~\ref{sec:results}.
Finally, in Sec.~\ref{sec:conclusions} we summarize our results and discuss possible future developments.

\section{Background}\label{sec:methods}

\subsection{Neural Networks}\label{sec:nn}

ANNs are a broad class of generic nonlinear universal approximators~\cite{goodfellow-book-2016}. 
Multilayer Perceptrons (MLPs) are the simplest ANN type: they are composed of neurons arranged in layers, 
$\mathcal{L}^{(i)}, i =1,\ldots, n$. 
Each layer consists of a learnable affine transformation, $\mathcal{A}^{(i)}$ (for the $i$-th layer), 
followed by a nonlinear activation function, $\sigma$, operating component-wise (i.e., independently on each output of 
the affine transformation). 
In formulas this reads
\begin{align}
  & \mathrm{y} = \mathcal{NN}(\mathrm{x}) = \mathcal{L}^{(n)} (\ \dots\  (\ \mathcal{L}^{(2)}\  (\mathcal{L}^{(1)} \mathrm{x})\ )\ )  
  & \text{\textbf{MLP}: composition of $n$ layers;} \label{eq:ann1} \\
  & \mathcal{L}^{(i)} \mathrm{x} = \sigma(\mathcal{A}^{(i)} \mathrm{x}), \:\:\: \forall\, i=1\dots n  
  & \text{\textbf{Layer}: affine transformation \& component-wise nonlinearity;} \label{eq:ann2}\\
  & \mathcal{A}^{(i)} \mathrm{x} = \mathbf{A}^{(i)} \mathrm{x} + \mathbf{b}^{(i)}, \:\:\: \forall\, i=1\dots n 
  & \text{\textbf{Affine transformation}: linear map \& translation;} \label{eq:ann3}
\end{align}
where $\mathrm{x}$ and $\mathrm{y}$ are, respectively, the vectors identifying the inputs and outputs of the network 
(or of the layers). Moreover, for each affine transformation $\mathcal{A}^{(i)} $, we denote with $\mathbf{A^{(i)}}$ 
its weight matrix (linear transformation), and $\mathbf{b^{(i)}}$ its bias vector (translation). 
In general, the output of each layer can be higher (or lower) dimensional than the input and/or the output dimension. 
These intermediate activations are also referred to as \textit{features}. 
This structure, composed of a juxtaposition of learnable affine operators and component-wise nonlinearities, 
is typically found in all modern architectures (e.g. convolutional, recurrent, attention blocks), 
and will be used also in the definition of our new architecture in the coming sections. 
Training MLPs consists of optimizing the parameters of the affine operation (weights and biases) in order 
to minimize some notion of difference between the prediction and the actual output, 
leveraging gradient descent and the backpropagation algorithm. 
For a detailed introduction, the reader is referred for example to~\cite{goodfellow-book-2016}.

\subsection{Lattice Boltzmann Method}\label{sec:lbm}
\begin{figure}[t]
  \centering
  \begin{subfigure}[b]{.4\linewidth}
      \centering
      \includegraphics[width=\linewidth]{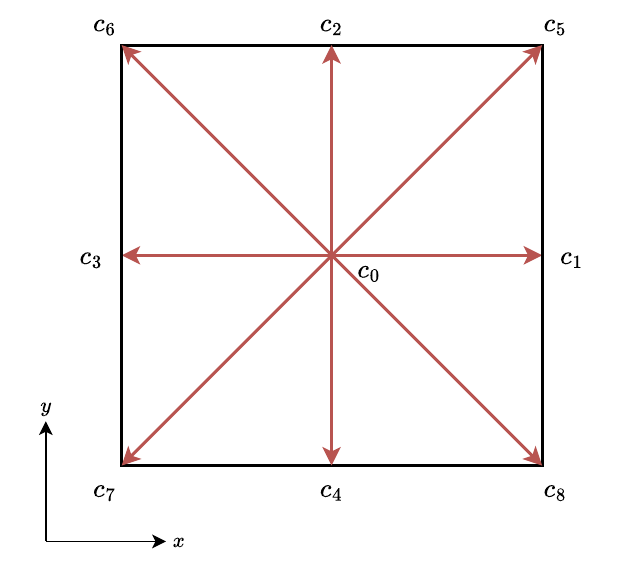}
      \caption{}
  \end{subfigure}
  \begin{subfigure}[b]{.4\linewidth}
      \centering
      \includegraphics[width=\linewidth]{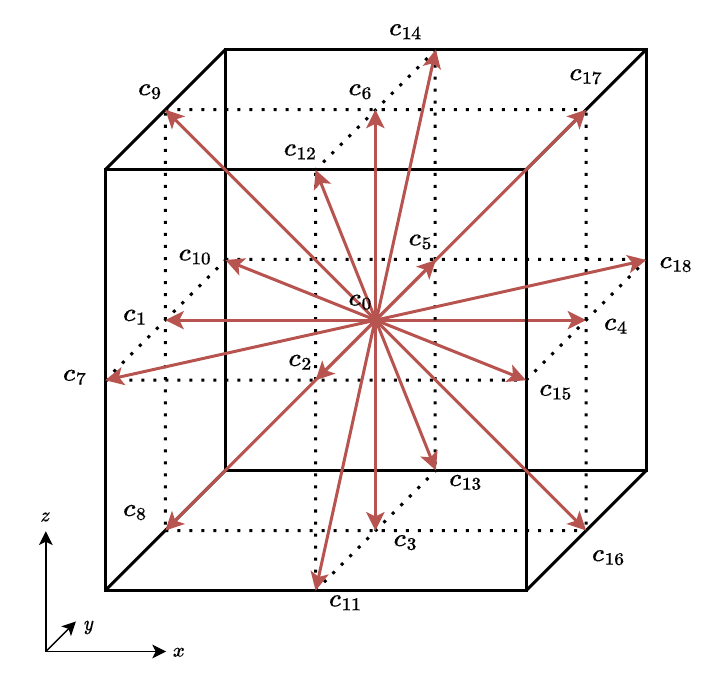}
      \caption{}
  \end{subfigure}
  \caption{Velocity stencils $\mathcal{V} = \{ \mathbf{c}_i \}$, with (a) the D2Q9 stencil and (b) the D3Q19 stencil.
         }\label{fig:stencils}
\end{figure}
In this section, we provide a brief introduction to the Lattice Boltzmann Method (LBM), which we use in this work 
as a prototype of lattice-based computational model.
LBM (see ~\cite{succi-book-2018,kruger-book-2017} for a detailed overview) is a class of numerical methods for the
simulation of fluid flows.
Unlike traditional computational fluid dynamics solvers which explicitly discretize the Navier-Stokes equations,
LBM stems from a discretization of the Boltzmann Equation, describing the state of the system in terms of
particle distribution functions $f(\mathbf{x},\mathbf{v}; t)$, giving at time $t$ the probability mass of particles 
in a small element of phase-space centered at position $\mathbf{x}$ with molecular velocity $\mathbf{v}$.  
A key ingredient of the method is the discretization of the momentum space~\cite{shan-jocs-2016}, which is performed employing 
a set of discrete velocity, $\mathcal{V} = \{ \mathbf{c}_i \}, i = 1, \dots Q$, referred to as \textit{velocity stencil} 
(see examples in Fig.~\ref{fig:stencils}), which coupling with a Gauss-type quadrature rule ensure that the hydrodynamic
moments of interest (e.g. density $\rho$ and velocity $\textbf{u}$) can be calculated via discrete sums 
on lattice populations $f_i(\textbf{x}, \textbf{c}_i)$, i.e. the the discrete counterpart of the particle distribution
functions: 
\begin{equation}
  \rho = \sum_i f_i , \quad \rho \mathbf{u} = \sum_i f_i \mathbf{c}_i \quad .
\end{equation}
The time evolution of the system is then dictated by the Lattice Boltzmann Equation, which can be evolved in
time according to the collide and stream paradigm:
\begin{align}
  & f_i^{\rm post}(\mathbf{x}, t          ) = f_i^{\rm pre }(\mathbf{x},t) + \Omega_i(\mathbf{x}, t)   & \text{\textit{collide}} \label{eq:lbmcoll}\\
  & f_i^{\rm pre }(\mathbf{x}, t+ \Delta t) = f_i^{\rm post}(\mathbf{x} - \mathbf{c}_{i} \Delta t , t) & \text{\textit{stream}}. \label{eq:lbmstream}  
\end{align}
A standard choice for modeling the collision operator $\Omega_i$ is the Bhatnagar-Gross-Krook (BGK)~\cite{bhatnagar-pr-1954} collision operator, 
consisting in a relaxation towards the local thermodynamic equilibrium
\begin{equation}\label{eq:bgk-collision}
  \Omega_i(\mathbf{x},t) = \frac{1}{\tau} \left(f_i^{\rm eq}(\mathbf{x},t)-f^{\rm pre}_i(\mathbf{x},t) \right),
\end{equation}
with relaxation rate $\tau$, and $f^{\rm pre}_i$ the discrete local equilibrium, for which we consider
a second-order expansion in Hermite polynomials of the Maxwell-Boltzmann distribution:
\begin{equation}\label{eq:feq}
        f^{\rm{eq}}_{i}(\rho(\mathbf{x},t), \mathbf{u}(\mathbf{x},t))   = \,  w_i \rho 
        \left( 
               1 + \frac{ \mathbf{u} \cdot \mathbf{c}_{i}}{c_s^2} 
                 + \frac{(\mathbf{u} \cdot \mathbf{c}_{i})^2 - (c_s |\mathbf{u}|)^2 }{2c_s^4}
       \right),
\end{equation}
with $c_s = 1 / \sqrt{3}$ the sound speed in the lattice, and $w_i$ a lattice-dependent set of weighting factors.
In this work we consider, the D2Q9 stencil for the 2D case and the D3Q19 velocity stencil for the 3D case, as shown in Fig.~\ref{fig:stencils} and reported in Appendix~\ref{sec:appendix-stenc}.

\subsection{Definition of the Learning problem: surrogate collision operators for the Lattice Boltzmann Method}\label{sec:problem-def}
We review here the framework introduced in Ref.~\cite{corbetta-epje-2023} for learning collision operators for LBM.
In particular we consider the task of learning with a ANN the BGK collision operator  (Eq.~\ref{eq:bgk-collision}); 
we make use of this learning problem as a reference for benchmarking our LENN models.  
In formal terms, this reads:
\begin{align}
  \text{\textit{train a ANN model,}} \quad & \mathcal{NN}(\cdot) \quad \text{\textit{such that}} \nonumber \\
                                           & \mathcal{NN}(\mathbf{f}^{\rm pre}) = \tilde{\mathbf{f}}^{\rm post} \approx \mathbf{f}^{\rm post}.
\end{align}
Hence, we aim at model whose input are the (local in space-time) values of the pre-collision populations, i.e., the vector $\mathbf{f}^{\rm pre}$, and returns an accurate approximation of the post-collision population vector, $\mathbf{f}^{\rm post}$.
This setting allows to establish a fully controlled setup for numerical experiments, making it easier to evaluate and 
compare different ANN architectures. %

\noindent We operate in a supervised learning context, thus the training dataset is composed of pairs of populations:
\begin{equation}
  \{ (\mathbf{f}^{\rm pre}_{k}, \mathbf{f}^{\rm post}_{k}), k=1,2,\cdots,N \}.
\end{equation}
In Ref.~\cite{corbetta-epje-2023}, and similarly in the forthcoming, the dataset is synthetically generated using the following relations %
\begin{align}\label{eq:dset-def}
    &\mathbf{f}^{\rm pre} = \mathbf{f}^{\rm eq}(\rho, \mathbf{u}) + \mathbf{f}^{\rm neq} \\
    &\mathbf{f}^{\rm post}= \mathbf{f}^{\rm pre} + \Omega
\end{align}
with the equilibrium $\mathbf{f}^{\rm eq}$ computed using Eq.~\ref{eq:feq} with respect to randomly sampled macroscopic fields
($\rho$ and $\mathbf{u}$), and with the non-equilibrium part $\mathbf{f}^{\rm neq}$ also randomly sampled and constrained 
to not introduce additional mass and momentum. %
\subsection{Lattice symmetries and lattice-equivariant operators}
The Boltzmann collision operator, along with its discrete counterpart $\Omega_i(\mathbf{x},t)$, 
is symmetric, i.e. equivariant, with respect to several transformations, including, 
homogeneity, space-time rescaling, and isotropy, i.e. symmetry with respect to rotation and reflection of the space. 
The latter geometric symmetry, in the discrete case, coincides with the restrictions of the orthogonal group $O_n$ to 
square or cubic lattices cells. Practically, these subgroups contain all the isometries of the square or of the cube, 
i.e. the  actions that, respectively, map the square into itself (in 2D) or the cube into itself (in 3D). 
Approximating through neural networks collision operators equivariant with respect to this symmetry group is the key 
component of this work. 
\begin{figure}[t]
\centering
  \begin{subfigure}[b]{.8\linewidth}
  \centering
  \includegraphics[width=\linewidth]{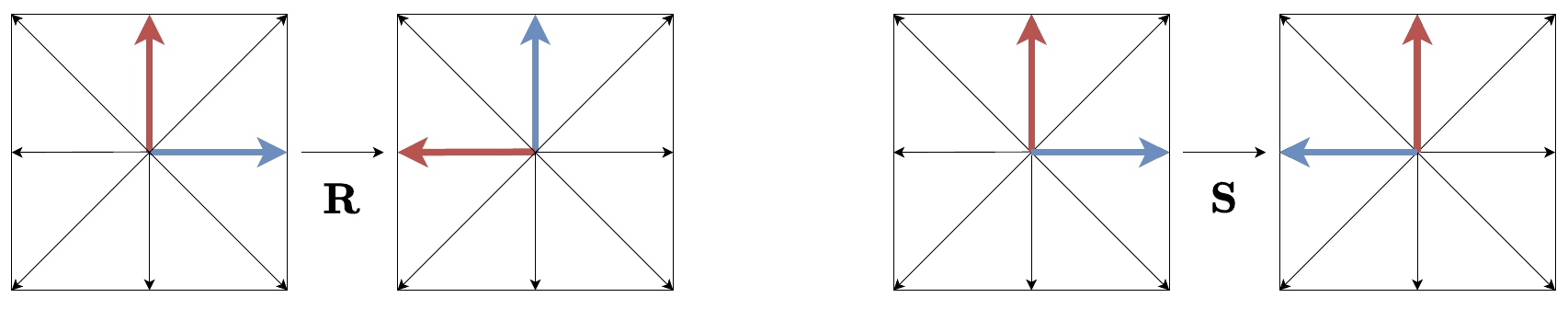}
  \caption{}
  \end{subfigure}
  \begin{subfigure}[b]{.8\linewidth}
  \centering
  \includegraphics[width=\linewidth]{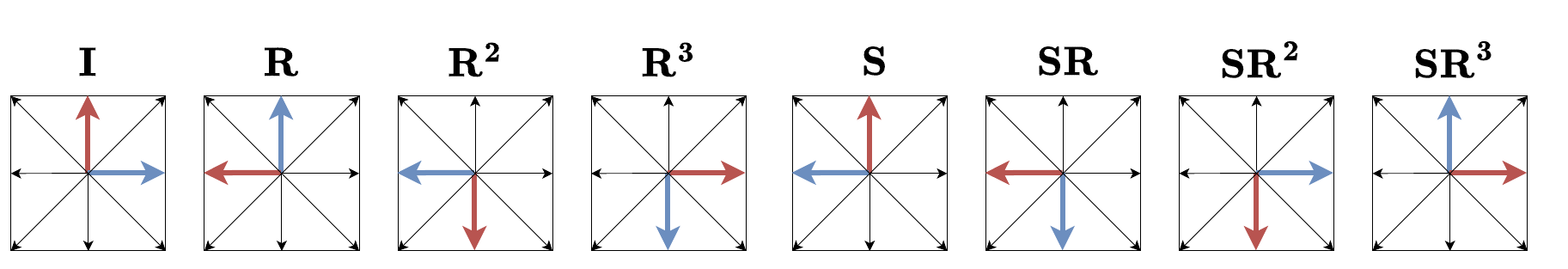}
  \caption{}
  \end{subfigure}
  \begin{subfigure}[b]{.8\linewidth}
  \centering
  \includegraphics[width=\linewidth]{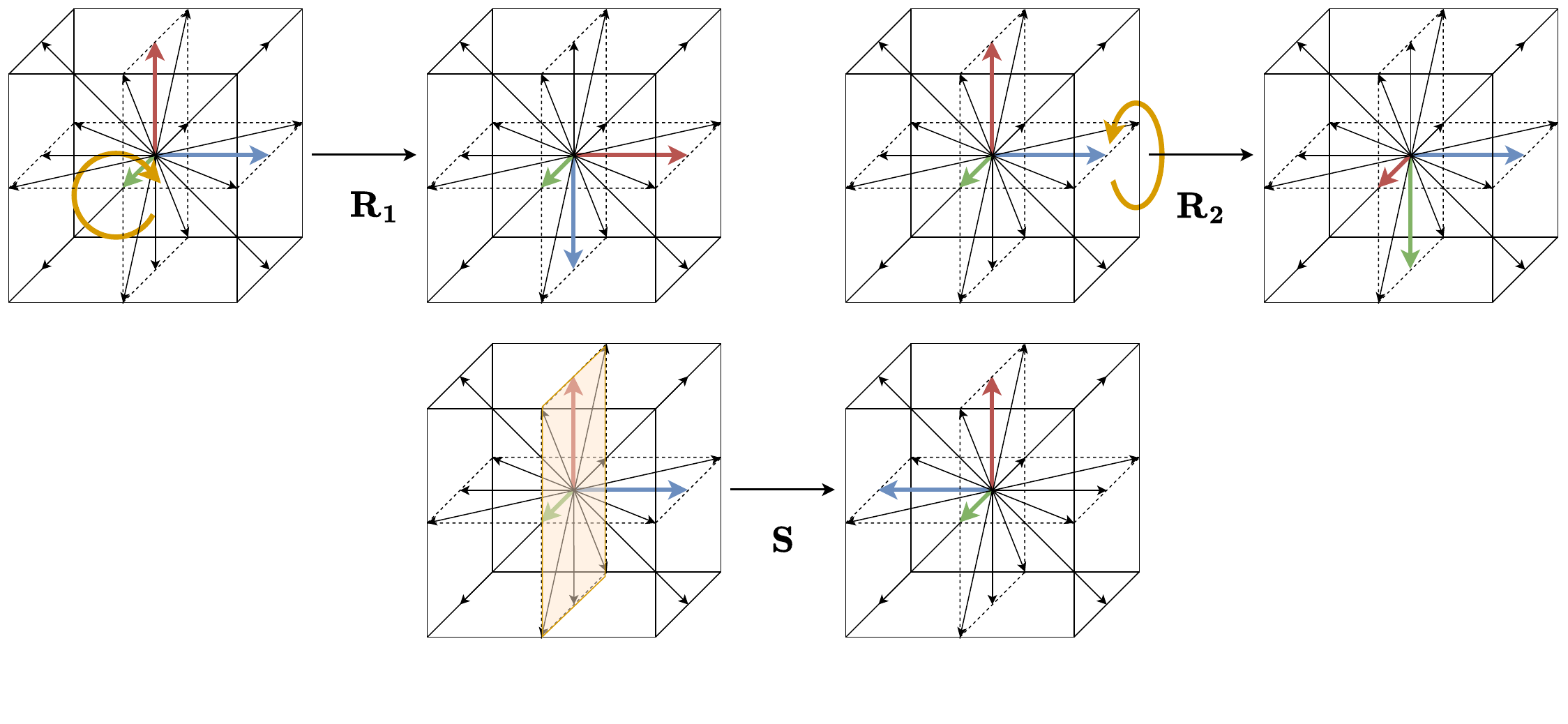}
  \caption{}
  \end{subfigure}
  \caption{(a,b) Visualization of the action of the dihedral symmetry group $D_4$ and the octahedral symmetry group $O_h$ as actions on the velocity stencils D2Q9 and D3Q19. 
           (a) The two $D_4$ generators: $\mathbf{R}$ (rotation by 90 degrees) and $\mathbf{S}$ (reflection). 
           (b) Representation of all symmetry group actions in $D_4$, that can be obtained as composition of the two generators (8 elements). 
           (c) The three $O_h$ generators: $\mathbf{R_1}$ 
           (rotation by 90 degrees around  the $x$ axis), $\mathbf{R_2}$ (rotation by 90 degrees around the $y$ axis), and $\mathbf{S}$ (reflection).
          }\label{fig:lattice-symmetries}
\end{figure}
Starting from the 2D case, the dihedral group, $G = D_4$, consists of all the isometries of the square, 
for a total of $8$ elements, i.e. four $90^{\circ}$ rotations and their combination with a 
reflection around one of the symmetry axis. 
Formally, we can express these transformations by combining together two \textit{generators},
the 90-degree rotation ($\mathbf{R}$, henceforth), and the reflection ($\mathbf{S}$), in formulas:
\begin{equation}
  D_4 = \generate{\mathbf{R}, \mathbf{S}} = 
       \{ \mathbf{I} = \mathbf{R}^0,
                       \mathbf{R},
                       \mathbf{R}^2,
                       \mathbf{R}^3,
                       \mathbf{S},
                       \mathbf{S}\mathbf{R},
                       \mathbf{S}\mathbf{R}^2,
                       \mathbf{S}\mathbf{R}^3 
       \} .
\end{equation}
In Fig.~\ref{fig:lattice-symmetries}a we report the generators of $D_4$, with in Fig.~\ref{fig:lattice-symmetries}b 
the corresponding elements associated to the application of each single transformation on a single lattice cell.
In 3D, the symmetry group to be considered is the octahedral group, $G=O_h$, corresponding to the symmetries of the cube. 
The generators of this group are two rotations along $x$ and $y$ axis, say $\mathbf{R_1}$ and $\mathbf{R_2}$, 
and a reflection across a symmetry plane, $\mathbf{S}$ (see Fig.~\ref{fig:lattice-symmetries}c). 
The combination of these $3$ generators give rise to a total of $48$ transformations, 
($24$ orientation preserving rotations plus $24$ non-orientation preserving rotations). 

Representation theory for finite groups establishes the existence of maps (representations) between 
the elements of a group and a finite set of non-degenerate square matrices of order $n$~\cite{artin-book}. 
More technically, each finite group is isomorphic to a certain closed subset of the general linear group of order $n$, $GL_n$,
comprising all non-degenerate linear operators $\mathbb{R}^n \rightarrow \mathbb{R}^n$. 
In our cases specialized to square lattice, the groups $G=D_4$ and $G=O_h$ can be represented as subsets, 
$\mathcal{M}(G) \subset P_q$, of the permutation matrices of $q$ elements, $P_q$. 
The elements of $\mathcal{M}(G)$ exchange, coherently with the group actions, the vertices/median-points of 
a lattice cell or the populations defined thereof ($\f = \{ f_0, f_1, \ldots , f_{q-1}\}$, 
cf. lattice stencil in Fig.~\ref{fig:stencils}). We report this correspondence schematically as: 
\begin{align}\label{eq:representation}
  \mbox{Lattice cell symmetries} &\leftrightarrow \mbox{Permutations of}\ \f = \{ f_0, f_1, \ldots, f_{q-1} \} \nonumber \\ 
  \text{Square}  \quad   G = D_4 &\leftrightarrow \mathcal{M}(D_4) = \generate{\mathbf{R},\mathbf{S}}\subset P_9\quad q=9  \\ %
  \text{Cube}    \quad   G = O_h &\leftrightarrow \mathcal{M}(O_h) = \generate{\mathbf{R_1},\mathbf{R_2},\mathbf{S}}\subset P_{19}\quad q=19 \nonumber %
\end{align}
In the 2D case, for instance, there thus exists two matrices representing the rotation by 90 degrees, 
$\mathbf{R}$, and the flip, $\mathbf{S}$, whose multiplication generates the whole group. 
These are, respectively,
\begin{equation}
  \mathbf{R} := 
  \begin{pmatrix}
    1 & 0 & 0 & 0 & 0 & 0 & 0 & 0 & 0 \\
    0 & 0 & 0 & 0 & 1 & 0 & 0 & 0 & 0 \\
    0 & 1 & 0 & 0 & 0 & 0 & 0 & 0 & 0 \\
    0 & 0 & 1 & 0 & 0 & 0 & 0 & 0 & 0 \\
    0 & 0 & 0 & 1 & 0 & 0 & 0 & 0 & 0 \\
    0 & 0 & 0 & 0 & 0 & 0 & 0 & 0 & 1 \\
    0 & 0 & 0 & 0 & 0 & 1 & 0 & 0 & 0 \\
    0 & 0 & 0 & 0 & 0 & 0 & 1 & 0 & 0 \\
    0 & 0 & 0 & 0 & 0 & 0 & 0 & 1 & 0 \\
  \end{pmatrix} \:\:\:\:\:\:\:\:
  \mathbf{S} := 
  \begin{pmatrix}
    1 & 0 & 0 & 0 & 0 & 0 & 0 & 0 & 0 \\
    0 & 1 & 0 & 0 & 0 & 0 & 0 & 0 & 0 \\
    0 & 0 & 0 & 0 & 1 & 0 & 0 & 0 & 0 \\
    0 & 0 & 0 & 1 & 0 & 0 & 0 & 0 & 0 \\
    0 & 0 & 1 & 0 & 0 & 0 & 0 & 0 & 0 \\
    0 & 0 & 0 & 0 & 0 & 0 & 0 & 0 & 1 \\
    0 & 0 & 0 & 0 & 0 & 0 & 0 & 1 & 0 \\
    0 & 0 & 0 & 0 & 0 & 0 & 1 & 0 & 0 \\
    0 & 0 & 0 & 0 & 0 & 1 & 0 & 0 & 0 \\
  \end{pmatrix}.
\end{equation}
Here and in what follows, we have used interchangeably the symbol for a group action and its matrix representation. 
We can now define the property of equivariance with respect to the group $G$ for a generic (non-linear) operator $\mathcal{F}$ as:
\begin{equation}\label{eq:equi}
  \mathcal{F} \colon \mathbb{R}^q \rightarrow \mathbb{R}^q\ \text{equivariant w.r.t. $G$} 
  \quad 
  \Leftrightarrow 
  \quad 
  \mathcal{F} (\mathbf{P} \f) = \mathbf{P} \mathcal{F}(\f), \:\:\: \forall\, \mathbf{P} \in \mathcal{M}(G),\ \forall\, \f \in \mathbb{R}^{q},
\end{equation}
i.e., in terms of operator composition ($\circ$):  
\begin{equation}\label{eq:equivariance}
  \mathcal{F} \colon \mathbb{R}^q \rightarrow \mathbb{R}^q\ \text{equivariant w.r.t. $G$} 
  \quad 
  \Leftrightarrow 
  \quad  
  \mathcal{F} \circ \mathbf{P} = \mathbf{P} \circ \mathcal{F}\quad \forall\, \mathbf{P} \in \mathcal{M}(G).
\end{equation}
We conclude this section reviewing two well-known properties, key in the derivation of our LENN layers.

\noindent \textbf{Property 1.} The composition of equivariant operators is equivariant, i.e.
          \begin{equation}\label{eq:nested-equivariance}
            \mathcal{F}_1,\ \mathcal{F}_2\quad  \mbox{equivariant w.r.t. $G$} 
            \Rightarrow 
            \mathcal{F}_1 \circ \mathcal{F}_2\quad \mbox{equivariant w.r.t. $G$};
          \end{equation}
          which can be proved by applying the definition in Eq.~\ref{eq:equi} twice consecutively.

\noindent \textbf{Property 2.} An operator is equivariant with respect to $G$, if and only if the equivariance 
          relation is satisfied for the group generators $g_1, g_2, \ldots$ (i.e., $G=\generate{g_1,g_2,\ldots}$). 
          In formulas, this reads:
          \begin{equation}\label{eq:eqgen}
            \mathcal{F} \circ \mathbf{P} = \mathbf{P}\mathcal{F}\quad \forall\, \mathbf{P} \in \mathcal{M}(G) 
            \quad  
            \Leftrightarrow  
            \quad 
            \mathcal{F} \circ \mathbf{P} = \mathbf{P}\mathcal{F}\quad \forall\, \mathbf{P} \in \genOfG.
          \end{equation}
This implies that in the 2D case it is sufficient to verify equivariance with respect to $\mathbf{R}$ and $\mathbf{S}$, 
and for the generators $\mathbf{R_1}$, $\mathbf{R_2}$ and $\mathbf{S}$ in 3D. 

The first property ensures that a network built of individually equivariant layers is itself equivariant, 
while the second reduces the number of independent constraints that will be enforced on each layer.

\subsection{Group Averaging}\label{sec:ga}
As discussed in Sec.~\ref{sec:intro}, an established method for enforcing equivariance with respect to a finite-sized 
symmetry group is Group Averaging (GAVG).
In short, given any non-equivariant operator $\mathcal{F}$, it is possible to define a new equivariant operator 
$\tilde{\mathcal{F}}$ as:
\begin{equation}\label{eq:groupavg}
  \tilde{\mathcal{F}}(\f) = \sum_{\mathbf{P} \in \mathcal{M}(G) } \mathbf{P}^{-1} \mathcal{F} (\mathbf{P} \f).
\end{equation}
Note that in the specific case of lattice symmetry, $\mathbf{P}$ is an orthogonal matrix (i.e. $\mathbf{P}^{-1} = \mathbf{P}^T$).
We remark that while this approach is straightforward to implement, and totally agnostic on the operator $\mathcal{F}$, 
its cost scales linearly with the number of symmetry actions, which as we discuss later on, can lead to significant
computational overheads.

\noindent In Appendix~\ref{sec:appendix-ga}, we provide an example of the application of Eq.~\ref{eq:groupavg} in the context 
of learning LBM collision operators in 2D.

\section{Lattice Equivariant Neural Networks}\label{sec:lenn}

In this section we introduce our novel Lattice-Equivariant Neural Network (LENN) architecture. 
Similarly to MLPs (Sec.~\ref{sec:nn}), we define LENNs as a composition of multiple layers, imposing the equivariance property Eq.~\ref{eq:equivariance} layer-by-layer.
From Eq.~\ref{eq:nested-equivariance} it follows immediately that an ANN consisting of individual equivariant layers 
will also be equivariant.
In what follows we therefore focus on the definition of a single layer of the network, for simplicity we will refer to such a layer with $\Layer$ instead of $\Layer^{(i)}$ (cf. Eq.~\ref{eq:ann1}).

\subsection{Lattice Equivariant layers}
We start by remarking that since lattice symmetries map into permutations of populations (cf. Eq.~\ref{eq:representation}), 
equivariant layers must be equivariant with respect to permutation (matrices) in $\mathcal{M}(G)$. 
Since the non-linearity $\sigma$ (Eq.~\ref{eq:ann2}) operates component-wise,  
it directly follows that $\sigma$ is also equivariant with respect to permutation 
(i.e., $\sigma \circ \mathbf{P} = \mathbf{P}\mathcal{\sigma}$). 
This is a key aspect in our derivation, since equivariance constraints need then to be applied only to affine operators 
$\mathcal{A}_i$ (Eq.~\ref{eq:ann3}). 
In order to define a single layer of the network we take the following four steps:
\begin{itemize}
    \item[\textbf{S1.}] Definition of equivariant affine operators acting on population vectors.
    \item[\textbf{S2.}] Generalization of population vectors into high-dimensional population features.
    \item[\textbf{S3.}] Definition of our LENN layers, i.e., generalized affine equivariant operators acting on population features.
    \item[\textbf{S4.}] Computationally efficient implementation.
\end{itemize}
\noindent The input of each layer will consist of population vectors, $\f$, or generalizations thereof, hence, as standard in the machine learning notation, will be  named $\x$.

\vspace{0.5cm}

\noindent \textbf{S1.} We consider an equivariant affine operator, $\mathcal{A}$, defined on population vectors 
           $\mathrm{x} \in \mathbb{R}^q$. By definition, $\mathcal{A}$ satisfies
          \begin{align}\label{eq:affine-1}
            & \mathcal{A}\colon \mathbb{R}^q \rightarrow \mathbb{R}^q \\
            & \mathcal{A}(\mathrm{x}) = \mathbf{A}\mathrm{x} + \mathbf{b}\ \quad \mathbf{A} \in \mathbb{R}^{q \times q}, \mathbf{b} \in \mathbb{R}^q \nonumber
          \end{align}
          where, as usual, $q$ is the number of lattice populations ($q = 9$ and $q=19$ in the example in Fig.~\ref{fig:stencils}). 
          By imposing the equivariance constraint Eq.~\ref{eq:equivariance} in Eq.~\ref{eq:affine-1}, 
          and considering the properties of generators (Eq.~\ref{eq:eqgen}), yields the following matrix-valued equation:
          \begin{equation}
            \mathbf{A} \mathbf{P} \x + \mathbf{b} = \mathbf{P} \mathbf{A} \x + \mathbf{P} \mathbf{b},
            \quad \forall\, \x\in \mathbb{R}^q,\ \forall\, \mathbf{P} \in \genOfG.
          \end{equation}
          The above equation is satisfied if and only if the weight matrix $\mathbf{A}$ and vector of bias $\mathbf{b}$ 
          respect the following under-determined set of constraints
          \begin{equation}\label{eq:equivariantA-constraint}
            \left\{
            \begin{aligned}
                \mathbf{A} \mathbf{P} &= \mathbf{P} \mathbf{A} \\
                \mathbf{b} &= \mathbf{P} \mathbf{b}
            \end{aligned}
            \right. \quad \forall\, \mathbf{P} \in \genOfG.
            \end{equation}
          For example, considering the D2Q9 stencil, $\mathbf{A}$ and $\mathbf{b}$ 
          satisfy Eq.~\ref{eq:equivariantA-constraint} if and only if they have the following form
          {
          \setlength{\arraycolsep}{3pt}
          \setcounter{MaxMatrixCols}{9}
          \begin{equation}\label{eq:mat-d2q9}
            \textbf{A}= \:\: \left( \:\:
            \begin{matrix}
              \aTwoD & \bTwoD & \bTwoD & \bTwoD & \bTwoD & \dTwoD & \dTwoD & \dTwoD & \dTwoD \\
              \cTwoD & \jTwoD & \kTwoD & \lTwoD & \kTwoD & \fTwoD & \gTwoD & \gTwoD & \fTwoD \\
              \cTwoD & \kTwoD & \jTwoD & \kTwoD & \lTwoD & \fTwoD & \fTwoD & \gTwoD & \gTwoD \\
              \cTwoD & \lTwoD & \kTwoD & \jTwoD & \kTwoD & \gTwoD & \fTwoD & \fTwoD & \gTwoD \\
              \cTwoD & \kTwoD & \lTwoD & \kTwoD & \jTwoD & \gTwoD & \gTwoD & \fTwoD & \fTwoD \\
              \eTwoD & \hTwoD & \hTwoD & \iTwoD & \iTwoD & \mTwoD & \nTwoD & \oTwoD & \nTwoD \\
              \eTwoD & \iTwoD & \hTwoD & \hTwoD & \iTwoD & \nTwoD & \mTwoD & \nTwoD & \oTwoD \\
              \eTwoD & \iTwoD & \iTwoD & \hTwoD & \hTwoD & \oTwoD & \nTwoD & \mTwoD & \nTwoD \\
              \eTwoD & \hTwoD & \iTwoD & \iTwoD & \hTwoD & \nTwoD & \oTwoD & \nTwoD & \mTwoD 
            \end{matrix} \:\: \right)
            \:\:\:\:\:\:\:\:\:\:\:\:\:\:\:\:
            \textbf{b} = \left( \:\: 
            \begin{matrix}
                \pTwoD \\
                \qTwoD \\
                \qTwoD \\
                \qTwoD \\
                \qTwoD \\
                \rTwoD \\
                \rTwoD \\
                \rTwoD \\
                \rTwoD \\
            \end{matrix} \:\: \right)
          \end{equation}
          }
          where we indicate with letters (and, equivalently with background colors) the independent free parameters 
          (i.e., weights). Hence, the equivariance constraint on the affine operator $\mathcal{A}$ 
          reduced the degrees of freedom from $q^2 + q =81+9=90$  to $\#_{A_q} + \#_{b_q} = 15+3=18$.
          We can define vectors containing the individual degrees of freedom, respectively
          \begin{equation}\label{eq:reduced}
            \tilde{\mathbf{A}} = (a,b,\ldots, o) \in \mathbb{R}^{\#_{A_q}} = \mathbb{R}^{15}, 
            \quad\quad 
            \tilde{\mathbf{b}} = (p,q,r) \in \mathbb{R}^{\#_{b_q}} = \mathbb{R}^3,
          \end{equation}
          which are in linear correspondence with $\mathbf{A}$ and $\mathbf{b}$, i.e. the following relations hold: 
          \begin{equation}\label{eq:delta-matrices}
            \mathbf{A}_{ij} = \sum_{k}\Lambda_{ijk} \tilde{\mathbf{A}}_k, 
            \quad\quad 
            \mathbf{b}_i = \sum_k \Delta_{ik} \tilde{\mathbf{b}}_k,
          \end{equation}
          for some 3D tensor $\Lambda_{ijk}$, and 2D tensor $\Delta_{ik}$. 
          Note that the components of tensors $\Lambda_{ijk}$ and $\Delta_{ik}$ can take only values $\in \{0, 1 \}$
          since these tensors replicate the degrees of freedom. 
          Here and whenever there is an ambiguity in matrix-multiplication operations we will highlight the component indices. 

          Following the same procedure for the octaedral symmetry group in 3D, we get employing the constraint in 
          Eq.~\ref{eq:equivariantA-constraint} the following results using as an example the D3Q19 stencil:
          \begin{equation}\label{eq:mat-d3q19}
            \textbf{A}= \left(
            \begin{xsmallmatrix}
              \aThreeD & \bThreeD & \bThreeD & \bThreeD & \bThreeD & \bThreeD & \bThreeD & \cThreeD & \cThreeD & \cThreeD & \cThreeD & \cThreeD & \cThreeD & \cThreeD & \cThreeD & \cThreeD & \cThreeD & \cThreeD & \cThreeD \\[-0.15em]
              \dThreeD & \fThreeD & \hThreeD & \hThreeD & \hThreeD & \hThreeD & \gThreeD & \iThreeD & \iThreeD & \iThreeD & \iThreeD & \kThreeD & \kThreeD & \kThreeD & \kThreeD & \jThreeD & \jThreeD & \jThreeD & \jThreeD \\[-0.15em]
              \dThreeD & \hThreeD & \fThreeD & \hThreeD & \hThreeD & \gThreeD & \hThreeD & \iThreeD & \kThreeD & \kThreeD & \jThreeD & \iThreeD & \iThreeD & \jThreeD & \jThreeD & \iThreeD & \kThreeD & \kThreeD & \jThreeD \\[-0.15em]
              \dThreeD & \hThreeD & \hThreeD & \fThreeD & \gThreeD & \hThreeD & \hThreeD & \kThreeD & \iThreeD & \jThreeD & \kThreeD & \iThreeD & \jThreeD & \iThreeD & \jThreeD & \kThreeD & \iThreeD & \jThreeD & \kThreeD \\[-0.15em]
              \dThreeD & \hThreeD & \hThreeD & \gThreeD & \fThreeD & \hThreeD & \hThreeD & \kThreeD & \jThreeD & \iThreeD & \kThreeD & \jThreeD & \iThreeD & \jThreeD & \iThreeD & \kThreeD & \jThreeD & \iThreeD & \kThreeD \\[-0.15em]
              \dThreeD & \hThreeD & \gThreeD & \hThreeD & \hThreeD & \fThreeD & \hThreeD & \jThreeD & \kThreeD & \kThreeD & \iThreeD & \jThreeD & \jThreeD & \iThreeD & \iThreeD & \jThreeD & \kThreeD & \kThreeD & \iThreeD \\[-0.15em]
              \dThreeD & \gThreeD & \hThreeD & \hThreeD & \hThreeD & \hThreeD & \fThreeD & \jThreeD & \jThreeD & \jThreeD & \jThreeD & \kThreeD & \kThreeD & \kThreeD & \kThreeD & \iThreeD & \iThreeD & \iThreeD & \iThreeD \\[-0.15em]
              \eThreeD & \lThreeD & \lThreeD & \nThreeD & \nThreeD & \mThreeD & \mThreeD & \oThreeD & \sThreeD & \sThreeD & \pThreeD & \sThreeD & \sThreeD & \rThreeD & \rThreeD & \pThreeD & \rThreeD & \rThreeD & \qThreeD \\[-0.15em]
              \eThreeD & \lThreeD & \nThreeD & \lThreeD & \mThreeD & \nThreeD & \mThreeD & \sThreeD & \oThreeD & \pThreeD & \sThreeD & \sThreeD & \rThreeD & \sThreeD & \rThreeD & \rThreeD & \pThreeD & \qThreeD & \rThreeD \\[-0.15em]
              \eThreeD & \lThreeD & \nThreeD & \mThreeD & \lThreeD & \nThreeD & \mThreeD & \sThreeD & \pThreeD & \oThreeD & \sThreeD & \rThreeD & \sThreeD & \rThreeD & \sThreeD & \rThreeD & \qThreeD & \pThreeD & \rThreeD \\[-0.15em]
              \eThreeD & \lThreeD & \mThreeD & \nThreeD & \nThreeD & \lThreeD & \mThreeD & \pThreeD & \sThreeD & \sThreeD & \oThreeD & \rThreeD & \rThreeD & \sThreeD & \sThreeD & \qThreeD & \rThreeD & \rThreeD & \pThreeD \\[-0.15em]
              \eThreeD & \nThreeD & \lThreeD & \lThreeD & \mThreeD & \mThreeD & \nThreeD & \sThreeD & \sThreeD & \rThreeD & \rThreeD & \oThreeD & \pThreeD & \pThreeD & \qThreeD & \sThreeD & \sThreeD & \rThreeD & \rThreeD \\[-0.15em]
              \eThreeD & \nThreeD & \lThreeD & \mThreeD & \lThreeD & \mThreeD & \nThreeD & \sThreeD & \rThreeD & \sThreeD & \rThreeD & \pThreeD & \oThreeD & \qThreeD & \pThreeD & \sThreeD & \rThreeD & \sThreeD & \rThreeD \\[-0.15em]
              \eThreeD & \nThreeD & \mThreeD & \lThreeD & \mThreeD & \lThreeD & \nThreeD & \rThreeD & \sThreeD & \rThreeD & \sThreeD & \pThreeD & \qThreeD & \oThreeD & \pThreeD & \rThreeD & \sThreeD & \rThreeD & \sThreeD \\[-0.15em]
              \eThreeD & \nThreeD & \mThreeD & \mThreeD & \lThreeD & \lThreeD & \nThreeD & \rThreeD & \rThreeD & \sThreeD & \sThreeD & \qThreeD & \pThreeD & \pThreeD & \oThreeD & \rThreeD & \rThreeD & \sThreeD & \sThreeD \\[-0.15em]
              \eThreeD & \mThreeD & \lThreeD & \nThreeD & \nThreeD & \mThreeD & \lThreeD & \pThreeD & \rThreeD & \rThreeD & \qThreeD & \sThreeD & \sThreeD & \rThreeD & \rThreeD & \oThreeD & \sThreeD & \sThreeD & \pThreeD \\[-0.15em]
              \eThreeD & \mThreeD & \nThreeD & \lThreeD & \mThreeD & \nThreeD & \lThreeD & \rThreeD & \pThreeD & \qThreeD & \rThreeD & \sThreeD & \rThreeD & \sThreeD & \rThreeD & \sThreeD & \oThreeD & \pThreeD & \sThreeD \\[-0.15em]
              \eThreeD & \mThreeD & \nThreeD & \mThreeD & \lThreeD & \nThreeD & \lThreeD & \rThreeD & \qThreeD & \pThreeD & \rThreeD & \rThreeD & \sThreeD & \rThreeD & \sThreeD & \sThreeD & \pThreeD & \oThreeD & \sThreeD \\[-0.15em]
              \eThreeD & \mThreeD & \mThreeD & \nThreeD & \nThreeD & \lThreeD & \lThreeD & \qThreeD & \rThreeD & \rThreeD & \pThreeD & \rThreeD & \rThreeD & \sThreeD & \sThreeD & \pThreeD & \sThreeD & \sThreeD & \oThreeD \\[-0.15em]
            \end{xsmallmatrix} \right) \:\:\:\:\:\:\:\:\:\:\:\:
            \textbf{b} = \left( 
            \begin{smallmatrix}
              \tThreeD \\[-0.15em]
              \uThreeD \\[-0.15em]
              \uThreeD \\[-0.15em]
              \uThreeD \\[-0.15em]
              \uThreeD \\[-0.15em]
              \uThreeD \\[-0.15em]
              \uThreeD \\[-0.15em]
              \vThreeD \\[-0.15em]
              \vThreeD \\[-0.15em]
              \vThreeD \\[-0.15em]
              \vThreeD \\[-0.15em]
              \vThreeD \\[-0.15em]
              \vThreeD \\[-0.15em]
              \vThreeD \\[-0.15em]
              \vThreeD \\[-0.15em]
              \vThreeD \\[-0.15em]
              \vThreeD \\[-0.15em]
              \vThreeD \\[-0.15em]
              \vThreeD \\[-0.15em]
            \end{smallmatrix} \right).
          \end{equation}
          In this case, the total amount of free parameters decreases from $361+19=380$ to $19+3=22$. 

\begin{figure}[t]
  \centering
  \begin{subfigure}[t]{.4\linewidth}
  \centering
  \includegraphics[width=\linewidth,trim={0 -1.5cm 0 0},clip]{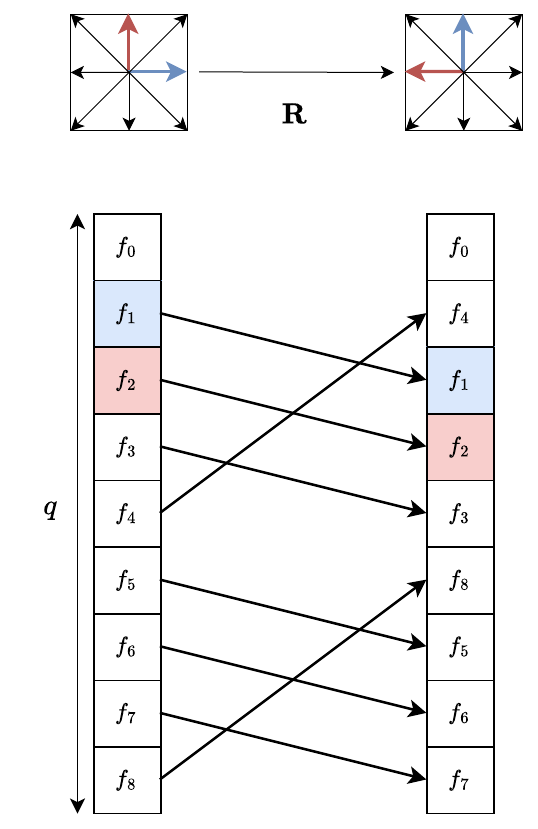}
  \caption{}
  \end{subfigure}
  \:\:\:
  \begin{subfigure}[t]{.48\linewidth}
  \centering
  \includegraphics[width=\linewidth]{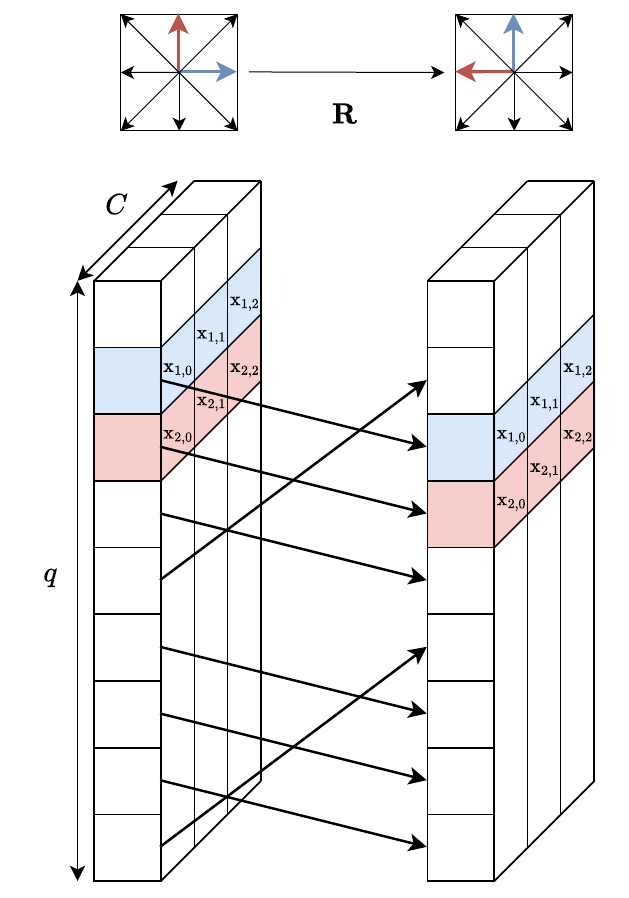}
  \caption{}
  \end{subfigure}
  \caption{ Visualization of the notion of generalized population features in LENN. (a) the rotation transformation $\mathbf{R}$ acting on a single population vector, $\f = \{ f_i \}$, 
            as permutation of populations (cf. Eq.~\ref{eq:representation}). 
            (b) the same action on the population features $\x = \{ \x _{i,c} \}$, where the permutation 
            now acts identically on the $C$ different channels (cf. Eq.~\ref{eq:genpop})}\label{fig:genpop}
\end{figure}

\noindent \textbf{S2.} The hidden (internal) layers of ANNs typically have inputs/outputs or variable dimensions, 
          generally larger than the input/output dimensions of the ANN itself. 
          This is such to allow for a large number of trainable parameters, in turn increasing the approximation 
          capacity of an ANN.  
          In order to replicate the same behaviour for our architecture, we introduce the notion of feature, 
          henceforth called \textit{population feature}, that generalizes population vectors having 
          the two following characteristics, as shown in Fig.~\ref{fig:genpop}:
          \begin{enumerate}
            \item population features are higher dimensional than population vectors and, particularly, they are matrix-valued:
            \begin{align}
              \text{\textit{population vector:}} \quad  \mathrm{x} =\f \in \R^q 
              \quad\quad 
              \rightarrow
              \quad\quad  
              \text{\textit{population feature:}} \quad  \mathrm{x}  \in \R^{q \times C}.
            \end{align}
            Drawing from the terminology of convolutional layers~\cite{lecun2015deep}, the number of additional components, $C$, 
            will be referred to as \textit{channel size}. Using this definition, the input and output of the LENN will 
            be regarded one-channel population features $\x \in \mathbb{R}^{q \times 1}$, 
            corresponding to the pre- and post-collision populations.  %
            \item the actions of the symmetry group, through permutation matrices $\mathbf{P}$, 
            are formally well-defined as they exchange coherently features connected to each population
            \begin{equation}\label{eq:genpop}
              \x \in \R^{q \times C} \rightarrow \Perm\x  \in \R^{q \times C}, \quad  \Perm\in \mathcal{M}(G).
            \end{equation}
          \end{enumerate}
          Increasing the numbers of intermediate channels in the network allows to arbitrarily increase 
          the number of learned features, yet, as we shall discuss in \textbf{S3.}, without violating equivariance. 

\begin{figure}[t]
\centering
\includegraphics[width=0.9\linewidth]{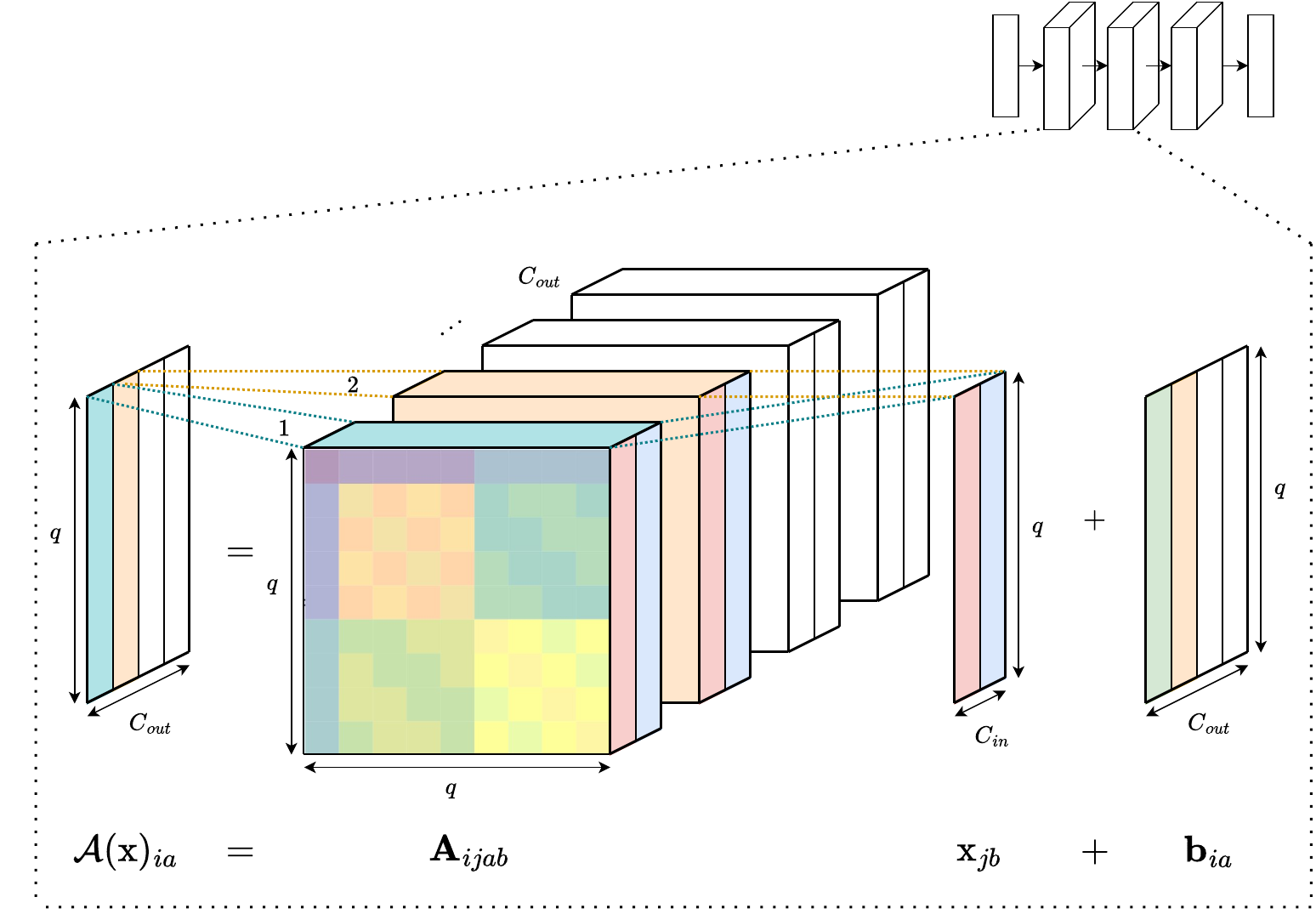}
\caption{Visualization of an equivariant layer, as outlined in Eq.~\ref{eq:equi}. 
         This layer operates with $C_{in}$ input channels and produces $C_{out}$ output channels. 
         The input $\x$ and output $\Affin(\x)$ have dimensions of $q \times C_{in}$ and $q \times C_{out}$, 
         respectively, where $q$ is the number of populations. 
         The linear matrix, with shape $q \times q \times C_{out} \times C_{in} $, 
         can be conceptualized as comprising $C_{out}$ blocks, each contributing to one of the output channels of $\mathcal{A}(\x)$. 
         Within each block, $C_{in}$ matrices of size $q \times q$ act on the channels of $\x$. 
         The submatrices of size $q \times q$ are constrained to conform to a specific shape, 
         as depicted in Eq.~\ref{eq:mat-d2q9} (for 2D) and~\ref{eq:mat-d3q19} (for 3D).
        }\label{fig:lenn}
\end{figure}

\noindent \textbf{S3. } Each layer of our LENN architecture operates an equivariant transformation which generalizes the 
         construction introduced in \textbf{S1} (Eq.~\ref{eq:equivariantA-constraint}) to population features (\textbf{S2}, Eq.~\ref{eq:genpop}) 
         having arbitrary number of channels. A sketch of our LENN layer is reported in Fig.~\ref{fig:lenn}. 
         Specifically, a LENN layer is a $G$-equivariant function such that
          \begin{align}
            & \Layer \colon \R^{q\times C_{in}} \rightarrow \R^{q\times C_{out}} \\
            & \x \mapsto \sigma (\Affin(\x)) \\
            & \Layer(\Perm \x) = \Perm \Layer(\x), \quad \forall \Perm \in \mathcal{M}(G).  
          \end{align}
          Due to the presence of channel axes we will write the next equations in components. Specifically, the affine part reads            
          \begin{align}
            & \x %
            \mapsto \quad \Affin(\x)_{ia} = \sum_{j,b} \mathbf{A}_{ijab} \x_{jb} + \mathbf{b}_{ia}, \\ %
            & \mathbf{A} \in \R^{q \times q \times C_{out} \times C_{in} } \quad \& \quad \mathbf{b} \in \R^{q \times C_{out}},
          \end{align}
          where $i,j$ are the output and input population indices and $a,b$ are the output and input channel indices.  
          Lattice equivariance of the layer depends only on the equivariance of the affine part. 
          This requires enforcing Eq.~\ref{eq:equivariance} over the population axis of the 4D tensor $\mathbf{A}$ and 2D tensor $\mathbf{b}$ as
          \begin{equation}
            \sum_j \mathbf{A}_{ijab} P_{jk} = \sum_j P_{ij} \mathbf{A}_{jkab} , 
            \quad\quad 
            \mathbf{b}_{ia} = \sum_j P_{ij} \mathbf{b}_{ja}, \:\:\: \forall a,b.
          \end{equation}
          Heuristically, to establish equivariance on the tensors $\mathbf{A}_{ijab}$ and $\mathbf{b}_{ia}$ it is 
          sufficient to require that each submatrix of shape $q \times q$ within $\mathbf{A}_{ijab}$, 
          and the subvectors of shape $q$ in $\mathbf{b}_{ia}$, 
          come with the form specified by Eq.~\ref{eq:mat-d2q9} for the 2D case and ~\ref{eq:mat-d3q19} for the 3D case.

\noindent \textbf{S4. } We conclude this section by outlining the implementation of our LENN layers, which can be 
          employed to train ANNs in established deep-learning packages (e.g. Tensorflow, PyTorch).
          From the observation  in Eq.~\ref{eq:reduced}, the number of degrees of freedom for tensors 
          $\mathbf{A}_{ijab}$ and $\mathbf{b}_{ia}$ is limited respectively to
          $\#_{A_q} \times C_{out} \times C_{in}$ and $\#_{b_q} \times C_{out}$. This makes it  possible to use the following 
          minimal memory representation
          \begin{equation}    
            \mathbf{\tilde A}_{kab}  \in \R^{\#_{A_q} \times C_{in} \times C_{out}}, 
            \quad \quad   
            \mathbf{\tilde b}_{ka} \in \R^{\#_{b_q} \times C_{out}}
          \end{equation}
          corresponding to the trainable weights available in each layer. 
          Combining the above with the tensors $\Lambda$ and $\Delta$ introduced in Eq.~\ref{eq:delta-matrices}, 
          it is possible to efficiently implement the affine part of a LENN layer composing matrix operations as
          \begin{equation}
            \Affin(\x)_{ia} %
                            = \sum_{j,b} \sum_k \Lambda_{ijk} \mathbf{\tilde A}_{kab} \x_{jb} + \sum_k \Delta_{ik} \mathbf{\tilde b}_{ka}. 
          \end{equation}

\section{Numerical Results}\label{sec:results}

In this section, we test and evaluate the LENN architecture introduced in Sec.~\ref{sec:methods}, considering the context of 
learning LBM collision operators (cf. Sec.~\ref{sec:problem-def}). 
We compare it against two other architectures: 1) A MLP with no equivariance constraints, 
and 2) a MLP employing the group-averaging method (cf. Sec.~\ref{sec:ga}). 

First, we detail the NN architectures and training approach, then, we evaluate performance and accuracy considering both 
a-priori and a-posteriori error metrics. The a-priori error analysis compares the ANNs considering their accuracy
on a (static) test set, of size $N_{\rm test}$, made of pre-generated pairs $\{(\f^{\rm pre},\f^{\rm post})\}$.  
In the a-posteriory analysis, we employ our ANNs in actual LBM simulations of both laminar (2D, 3D) and turbulent flows (3D), 
and estimate errors considering complete time dynamics in comparison with ground truth BGK evolution (cf. Sec.~\ref{sec:lbm}).

\subsection{Neural Networks Architecture and Training}\label{sec:res-pre}

All the architecture in our comparisons, MLP, GAVG, LENN, are meant to operate in the following context, 
in line with LBM collision operator learning problem introduced in Sec.~\ref{sec:problem-def}:
\begin{equation}\label{eq:actual-operator-learning-problem}
  \tilde{\mathbf{f}}^{\rm post} %
  = \mathbf{f}^{\rm pre} + \frac{1}{\tau} \Omega^{\mathcal{NN}}(\mathbf{f}^{\rm pre}),
\end{equation}
where our ANN appears as the term $\Omega^{\mathcal{NN}}$, which is here scaled by $\tau$ 
in order to make the learning problem independent of the choice of the relaxation time.
Additionally, following Ref.~\cite{corbetta-epje-2023}, we require that mass and momentum are conserved across the (learned) collision process. 
These conservation properties are expressed by the constraints
\begin{align}\label{eq:nn-conservation}
  \sum_i \Omega^{\mathcal{NN}}(\mathbf{f}^{\rm pre})_i                = 0           , \quad\quad\quad
  \sum_i \Omega^{\mathcal{NN}}(\mathbf{f}^{\rm pre})_i \mathbf{c}_{i} = \mathbf{0}  .
\end{align}
For instance, for the D2Q9 stencil, these constraints are satisfied by correcting the neural network output as
\begin{equation}    
 \Omega^{\mathcal{NN}}(\mathbf{f}^{\rm pre})_i 
 \leftarrow 
 \Omega^{\mathcal{NN}}(\mathbf{f}^{\rm pre})_i + \kappa_1 + \kappa_2 \mathbf{c}_{i,x} + \kappa_3 \mathbf{c}_{i,y},
 \label{eq:cons2}
\end{equation}
with
\begin{equation}  
  \kappa_1           = - \frac{1}{9} \sum_i \Omega^{NN}(\mathbf{f}^{\rm pre})_i, \quad\quad
  \kappa_{\{ 2,3\}}  = - \frac{1}{6} \sum_i \Omega^{NN}(\mathbf{f}^{\rm pre})_i \mathbf{c}_{i\{ x,y\}},
\end{equation}
For the D3Q19 stencil, it holds, instead:
\begin{equation}
 \Omega^{\mathcal{NN}}(\mathbf{f}^{\rm pre})_i 
 \leftarrow 
 \Omega^{\mathcal{NN}}(\mathbf{f}^{\rm pre})_i + \kappa_1 + \kappa_2 \mathbf{c}_{i,x} + \kappa_3 \mathbf{c}_{i,y} + \kappa_4 \mathbf{c}_{i,z}.
 \label{eq:cons3}
\end{equation}
with
\begin{equation}  
  \kappa_1  = - \frac{1}{19} \sum_i \Omega^{\mathcal{NN}}(\mathbf{f}^{\rm pre})_i, \quad\quad
  \kappa_{\{ 2,3,4\}}  = - \frac{1}{10} \sum_i \Omega^{\mathcal{NN}}(\mathbf{f}^{\rm pre})_i \mathbf{c}_{i,\{x,y,z\}}.
\end{equation}
Note that, since the constraints Eq.~\ref{eq:cons2},~\ref{eq:cons3} are imposed isotropically, it can be shown they do not compromise equivariance.

To establish a fair comparison between the different architectures, we tune the number and size of layers (for the MLPs), 
i.e., the number of channels (for LENN), to have a fixed number of free trainable parameters across the different architectures. 
The details of the architectures are reported in Table~\ref{table:freepars}.
\renewcommand{\arraystretch}{1}
\begin{table}[]
  \begin{center}   
  \begin{tabular}{ |c|c|c|c|c|c| } 
    \hline
                         &\textbf{Model} & \textbf{Architecture}  & \textbf{\# Free parameters} & \textbf{ \# Total parameters}  & \textbf{Ratio}  \\ 
    \hline \hline
    \multirow{ 3}{*}{2D} & MLP             & [9,13,13,13,9] & $2,340$                     & $2,340$                        & $1$             \\ 
    \cline{2-6}
                          & GAVG            & [9,13,13,13,9] & $2,340$                     & $18,720 $                      & $8$             \\ 
    \cline{2-6}
                          & \textbf{LENN}   & [1,10,8,8,1] & $2,550$                     & $13,770$                      & $5.4$           \\ 
    \hline \hline
     \multirow{ 3}{*}{3D} & MLP             & [19,50,50,50,19] & $6,900$                     & $6,900$                        & $1$             \\ 
    \cline{2-6}
                          & GAVG            & [19,50,50,50,19] & $6,900$                     & $331,200$                      & $48$            \\ 
    \cline{2-6}
                          & \textbf{LENN }  & [1,30,30,30,1] & $6,916$                     & $131,404$                      & $19$            \\ 
    \hline
  \end{tabular}
  \end{center}
  \caption{Details for the architectures of for the different models considered. For MLP and GAVG, the numbers reported in the "Architecture" column refer to the number of neurons in each layer, while for LENNs they refer to the number of channels for each layer. The free parameters are the independent trainable weights of the networks. The ratios between the total parameters and free parameters parameters are reported in the last column. For the group averaging approach the ratio values (8 for 2D; 48 for 3D) follow the order of the symmetry group considered; in the case of our LENN the ratios are connected to the reduced degrees of freedom of the equivariant layers (cf. Eq.~\ref{eq:reduced}). The reduced number of total parameters for the LENN with respect to GAVG imply a reduced computational cost, without degradation of the expressive power of the models, as shown in Fig.~\ref{fig:aposteriori}.
          }\label{table:freepars}
\end{table}
Following the problem definition given in Sec.~\ref{sec:problem-def}, the training and testing dataset are generated considering a value of $\rho=1$ and $u_x$, $u_y$ and (for 3D only) $u_z$ sampled uniformly in the interval $[-0.05,0.05]$. %
The non-equilibrium part $\mathbf{f}^{\rm neq}$ (cf. Eq.~\ref{eq:dset-def}) is sampled from a Normal distribution $\mathcal{N}(\mu=0, \sigma=10^{-3}$. The training and testing dataset sizes are respectively $N_{\rm train} = 10^5$ and $N_{\rm test} = 10^4$.

The training of all models is performed using stochastic gradient descent with an Adam optimizer~\cite{kingma-arxiv-2014}, with batch size $B = 128$, with initial learning rate $1 \times 10^{-3} $, and for a total of 200 epochs, which ensures convergence of the training process.
We use the Mean Square Error (MSE), computed on the collision term (Eq.~\ref{eq:bgk-collision}), as training loss:
\begin{equation}\label{eq:mse}
  \mathbf{L_{MSE}} = \sum_{k=0}^{B} || \Omega_k - \Omega_k^{\mathcal{NN}} ||^2.
\end{equation}
To accelerate the first phase of training, the ground truth collision term $\Omega$ is normalized to zero mean and unitary variance, using the same scaling factor for all the populations, The conservation of mass and momentum is imposed by first de-normalizing the output, then applying Eq.~\ref{eq:cons2}~ (2D) or \ref{eq:cons3} (3D), and re-normalizing to compute the loss. 
Empirically, we have observed the absolute error metric in Eq.~\ref{eq:mse} to be the most effective choice for the training of the Neural Network; note, however, that later on in Section~\ref{sec:results} all models' performance will be evaluated using relative error measures.
Finally, in order to asses the robustness and reproducibility of the training, each architecture is trained on the 
same synthetic dataset $20$ times, with each instance starting from an independent random initialization of weights and bias parameters. 

\subsection{A-priori error}
\begin{figure}[htb]
  \centering
  \begin{overpic}[width=\linewidth]{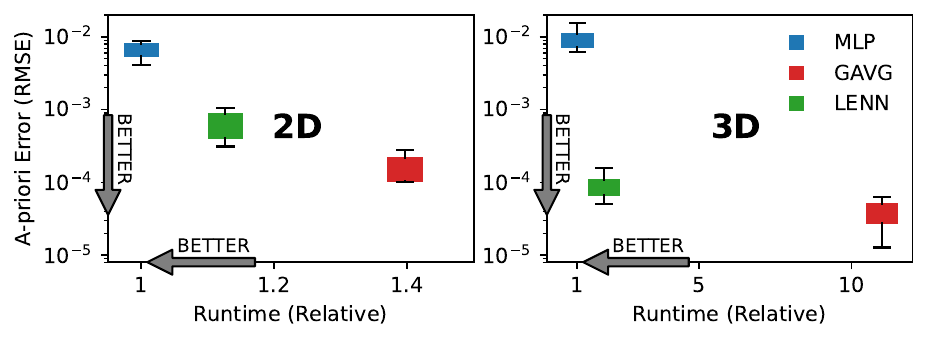}
    \put(28,  -1){\textbf{ (a) }}
    \put(75,  -1){\textbf{ (b) }}
  \end{overpic}
  \caption{Comparison of a-priori error versus relative execution time for the non-equivariant MLP (blue), MLP with group averaging (GAVG) (red) and LENN (green), for 2D (a) and 3D (b). The a-priori error is measured as average Relative Mean Square Error (RMSE) on the post-collision populations on the test set (Eq.~\ref{eq:apriori}). The execution times are evaluated at testing, and are normalized on the non-equivariant MLP value (the faster time). The barplots are computed considering 20 different architectures trained independently. The equivariant models (GAVG and LENN) show improved accuracy with respect to the non-equivariant MLP. GAVG shows a slight improved accuracy over LENN, especially in the 2D setting, not observed in the a-posteriori error (Fig.~\ref{fig:aposteriori}). However, LENN is considerably faster than GAVG, especially in the 3D setting.
  }\label{fig:apriori}
\end{figure}
We start considering an a-priori analysis, where we evaluate the performance of the different ANNs architectures comparing the average Relative Mean Square Error (RMSE) between the prediction of the model and the ground truth post-collision population, within a predefined test dataset:
\begin{equation}\label{eq:apriori}
  \mathbf{L^{a-priori}_{RMSE}} = \sum_{k=0}^{N_{\rm test}} \frac{|| \tilde{\textbf{f}}_k^{\rm post} - \textbf{f}_k^{\rm post} ||^2}{||\textbf{f}_k^{\rm post}||^2}
\end{equation}
In Fig.~\ref{fig:apriori} we report results both in two and three dimensions, where the accuracy of each network architecture, shown with boxplots summarizing the accuracy of 20 independently trained instances, is plotted versus the execution time.  Times are computed at testing, and scaled with respect to the non-equivariant MLP, which serves as the benchmark for speed.
Both GAVG and LENNs exhibit improved a-priori accuracy with respect to the non-equivariant MLP, albeit with a slightly better accuracy observed for the GAVG. 
On the other hand, LENNs outperform GAVG models significantly in terms of evaluation time. This holds particularly in the 3D setting where the evaluation time improves by one order of magnitude, highlighting the improved scalability of LENNs with respect to GAVG models. 

\subsection{A-posteriori error}\label{sec:res-apo}
We now consider a error analysis in the application of the different ANNs models for simulating the time dynamics of fluid flows in 2D and 3D.
Starting from given initial conditions, we compare the predicted dynamics agaist the ground truth dynamics obtained from simulations employing the BGK collision operator. All the test cases concern freely decaying flow conditions, i.e. no forcing, on square $(L^2)$ and cubical domains $(L^3)$ for the 2D and 3D case respectively, with $L=32$, following the problem setting defined in~\cite{corbetta-epje-2023}.
\begin{figure}[htb]
  \centering
  \begin{subfigure}[b]{\linewidth}
  \centering
  \includegraphics[width=\linewidth]{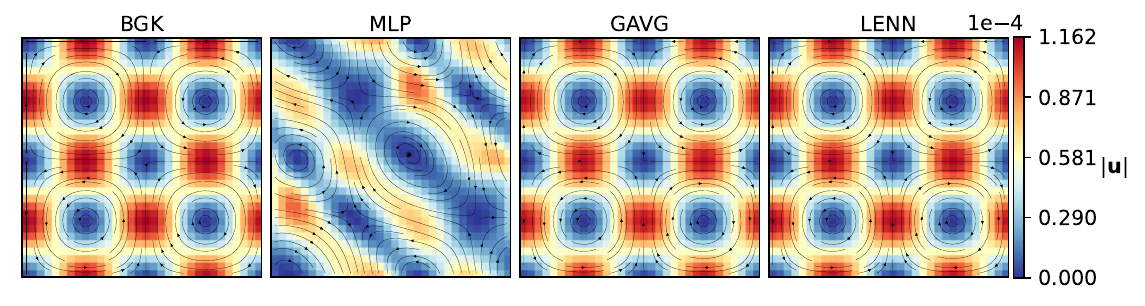}
  \caption{} 
  \end{subfigure}
  \begin{overpic}[width=\columnwidth]{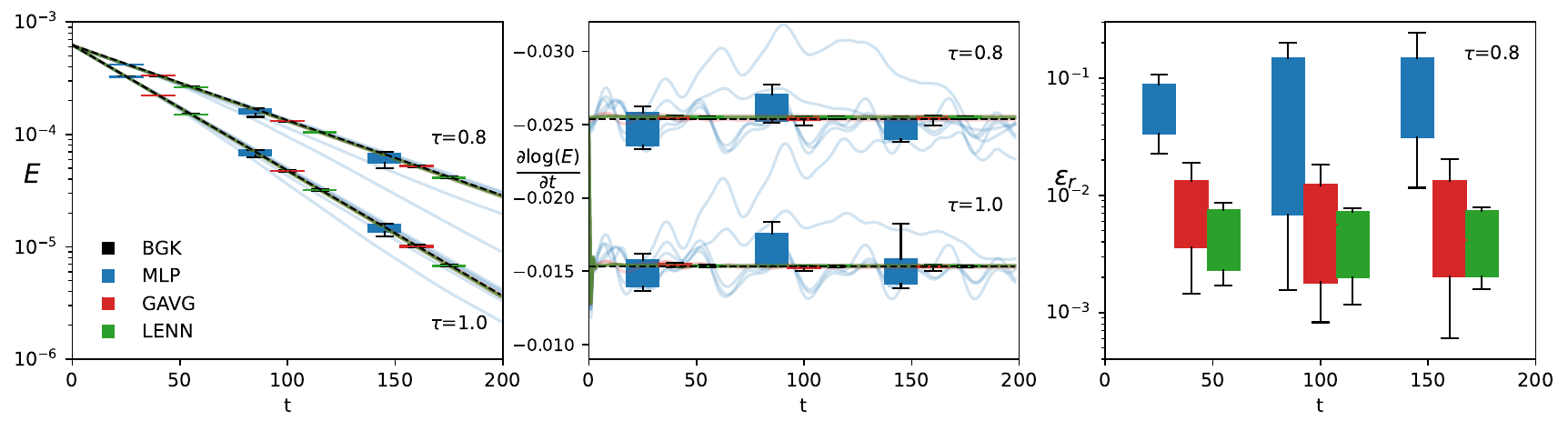}
    \put(15,  -1){\textbf{ (b) }}
    \put(49,  -1){\textbf{ (c) }}
    \put(83,  -1){\textbf{ (d) }}
  \end{overpic}  
  \caption{Results for the 2D Taylor-Green Vortex Decay test case for two different relaxation times $\tau=0.8$, $\tau=1$. 
           Panel (a) shows a qualitative comparison of the velocity magnitude $|\mathbf{u}|$, with velocity streamlines, 
           for the non-equivariant MLP, MLP with group averaging (GAVG) and LENN.
           Panels (b)-(d) illustrate a quantitative comparison for the total energy decay. 
           In panel (b), we show the total energy over time in a logarithmic scale. 
           In panel (c), we show the logarithmic derivative of velocity, representing the local exponential decay rate. 
           In panel (d), we show the relative absolute difference between the predicted local decay rate and 
           the ground truth decay rate, $\epsilon_r$ (Eq.~\ref{eq:errorrate}), for the different models. 
           The barplots are computed considering 20 different architectures trained independently, 
           while for (a) and (b), the individual lines correspond to the evolution for 5 such architectures. 
           The non-equivariant MLP is shown to disrupt the symmetric structure of the flow and to yield inconsistent 
           training performances on the energy decay, while the two equivariant models display good performances in 
           terms of flow structure and predicted decay rate.
           }
  \label{fig:taylor-2d}
\end{figure}

We start from the Taylor-Green vortex flow in 2D, with the following initial conditions:
\begin{align}
  & u_x(x,y; t_0) =  A \sin\left( k x\right) \cos\left( k y\right), \ \quad \ \quad 
   u_y(x,y; t_0) = - A \cos\left( k x\right) \sin\left( k y\right).
\end{align}
with $A = 0.05$, $k = 2 \pi / L$ and $x, y \in [0, L)$.
This is a well known example of unsteady decaying flow, with an exact solution for the (incompressible) Navier-Stokes equations.
Considering the kinetic energy energy $E = \frac{1}{2} \sum |u|^2$, its time evolution is characterized by an exponential decay with a fixed rate given by: 
\begin{equation}
    E(t) = E(0) \exp{ \left( -\nu k^2 t \right) },
\end{equation}
where $\nu$ is the viscosity of the fluid, which in LBM is put in relationship with the relaxation time parameter via $\nu = (\tau - \frac{\Delta t}{2} ) c_s^2$. To verify the generalization capabilities of the model with respect to $\tau$,  we evaluate the different models considering two different relaxation parameters values: $\tau=0.8$ and $\tau=1$.
\begin{figure}[t]
  \centering
  \begin{overpic}[width=\linewidth]{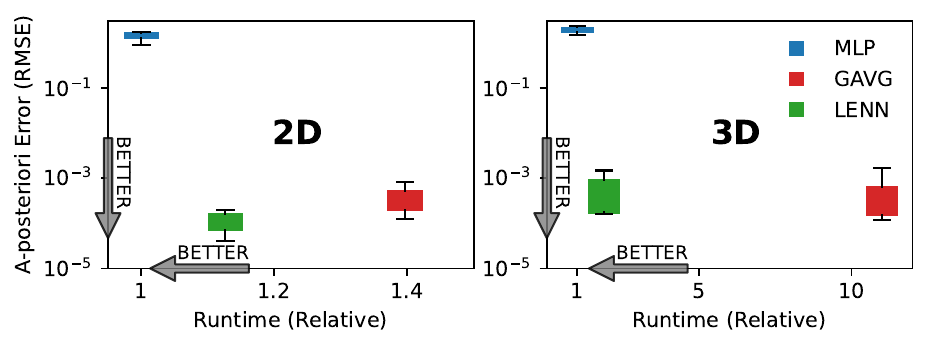}
    \put(28,  -1){\textbf{ (a) }}
    \put(75,  -1){\textbf{ (b) }}
  \end{overpic}
  \caption{Comparison of a-posteriori error versus Relative execution time for the non-equivariant MLP (blue), 
           MLP with group averaging, GAVG (red), and LENN (green), for 2D (a) and 3D (b). 
           The a-posteriori error is measured as average Relative Mean Square Error (RMSE) on the velocity (Eq.~\ref{eq:aposteriori}). 
           The execution times are evaluated at testing and are normalized on the non-equivariant MLP value (the faster time). 
           The barplots are computed considering 20 different architectures trained independently. 
           The equivariant models (GAVG and LENN) show improved accuracy with respect to the non-equivariant MLP. 
           However, LENN is considerably faster than GAVG, especially in the 3D setting.}\label{fig:aposteriori}
\end{figure}

The results are reported in Fig.~\ref{fig:taylor-2d}. Panel (a) shows the value of the velocity magnitude $|u|$ for the ground truth BGK solution and one particular realization for each of the models after $200$ timesteps, along with streamlines. While the two equivariant models, GAVGs and LENNs, yield results that are qualitatively identical to the BGK solution, the non-equivariant MLP instead fails, disrupting the symmetric structure of the flow and showcasing the need for equivariance in the model. 
On a more quantitative ground, panels (b)-(d) of Fig.~\ref{fig:taylor-2d} report the total energy, $E$, and derived quantities. Panel (b) shows the value of the $E$ for the first $200$ timesteps: while GAVG and LENNs are able to consistently capture the exponential decay for both values of the viscosity, the standard MLP fails. This is more evident in panel (c), which reports the value of the logarithmic derivative of $E$, $\frac{\partial log(E)}{\partial t}$ in the interval $t \in [50,200]$. In such plot an exponential decay of $E$ is represented by a constant value for the logarithmic derivative, different for the two $\tau$ values. This is indeed the case for the two equivariant models,  GAVG and LENN, while significant oscillations around the ground-truth can be observed for the MLP, as shown by the different realizations and the boxplots. Finally, in panel (d), we 
take the comparison to even finer-level scale, showing the relative absolute error between the predicted logarithmic derivative and the ground truth decay rate:
\begin{equation}
    \epsilon_r = \frac{\left\vert \frac{\partial log(E)}{\partial t} - \frac{\partial log(E_{BGK})}{\partial t} \right\vert}{\left\vert\frac{\partial log(E_{BGK})}{\partial t}\right\vert} .
\end{equation}\label{eq:errorrate}
In terms of accuracy, on average, LENNs manage to reach results comparable with GAVGs, at times better, yet retaining the speedup in computational times observed in the a-priori error analysis.
Similar considerations apply when repeating this laminar test case in 3D; for completeness the results are reported in the Appendix Sec.~\ref{sec:appendix-res}. 
\begin{figure}[t]
  \centering
 \begin{subfigure}[b]{\linewidth}
  \centering
  \includegraphics[width=\linewidth]{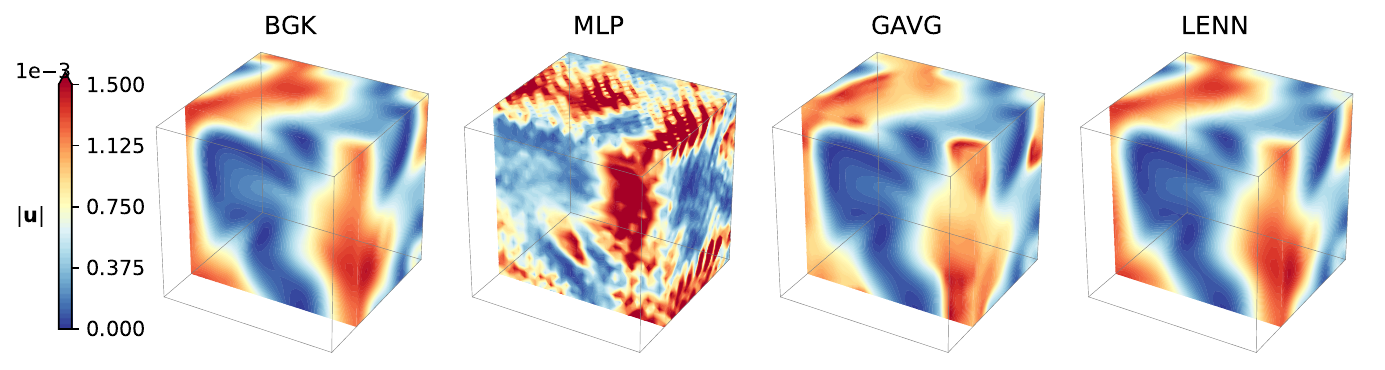}
  \caption{} 
  \end{subfigure}
  \begin{overpic}[width=\columnwidth]{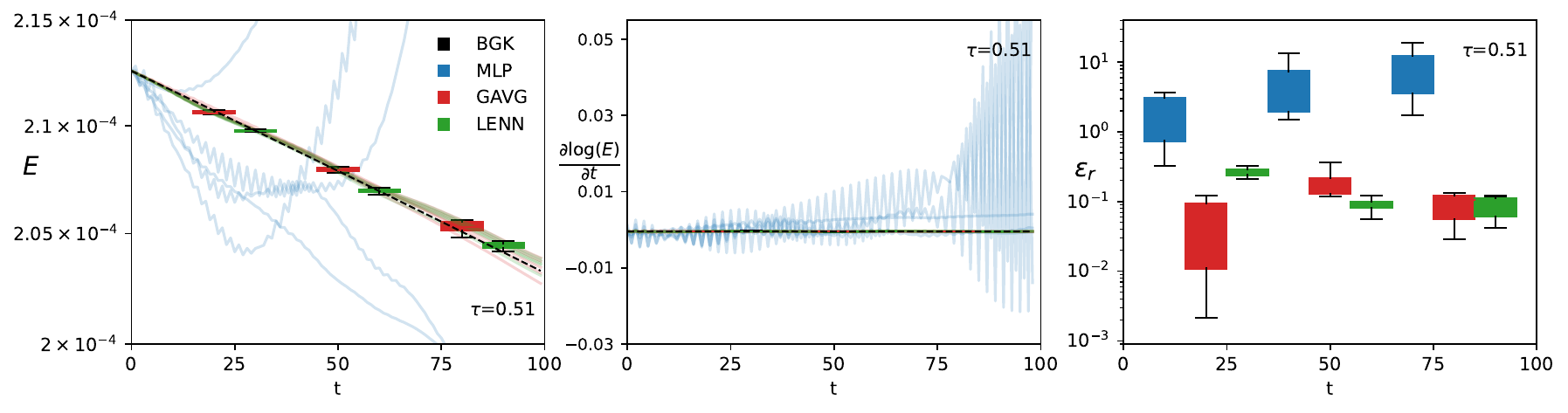}
    \put(15,  -1){\textbf{ (b) }}
    \put(49,  -1){\textbf{ (c) }}
    \put(83,  -1){\textbf{ (d) }}
  \end{overpic}    
  \caption{Results for the 3D Turbulent Decay test case for relaxation time $\tau=0.51$. Panel (a) shows a qualitative comparison of the velocity magnitude $|\mathbf{u}|$ for the non-equivariant MLP, MLP with group averaging (GAVG) and LENN.
  Panels (b)-(d) illustrate quantitative comparison for the total energy decay. On panel (b), we show the total energy over time in a logarithmic scale. On panel (c), we show the logarithmic derivative of velocity, representing the local exponential decay rate. On panel (d), we show the relative absolute difference between the predicted local decay rate and the ground truth decay rate, $\epsilon_r$ (Eq.~\ref{eq:errorrate}), for the different models. The barplots are computed considering 20 different architectures trained independently, while for (a) and (b), the individual lines correspond to the evolution for 5 such architectures. The non-equivariant MLP yields unstable dynamics, showcasing the need for equivariance in the model architecture to learn correct decay in the turbulent setting.}
  \label{fig:turbulent-3d}
\end{figure}

Considering the results obtained in the 2D and 3D laminar setting, we report in Fig.~\ref{fig:aposteriori} the a-posteriori error of the different models as a function of the models execution times (cf. a-priori analogous in Fig~\ref{fig:apriori}). The a-posteriori error is computed as Mean Square Error on the velocity, integrated in space and averaged in time on the first $200$ timesteps: 
\begin{equation}\label{eq:aposteriori}
  \mathbf{L^{\rm a-posteriori}_{MSE}} = \frac{1}{T} \sum_{t=1}^{T} \int \frac{|| \tilde{\mathbf{u}}(\mathbf{x},t) - \mathbf{u}(\mathbf{x},t) ||^2}{ ||\mathbf{u}(\mathbf{x},t) ||^2} d\mathbf{x},
\end{equation}
with $T=200$, $\tilde{\mathbf{u}}(\mathbf{x},t)$ the value of the velocity obtained from the different models (MLP, GAVG and LENN) and 
 $\mathbf{u}(\mathbf{x},t)$ the ground truth velocity. The results are shown in Fig.~\ref{fig:aposteriori}, further supporting the considerations made in the a-priori error (cf. Fig.~\ref{fig:genpop}). Additionally, we note that, in the case of the a-posteriori error, the accuracy of the two equivariant models, LENNs and GAVGs, is comparable.
\begin{figure}[t]
  \centering
  \includegraphics[width=.8\linewidth]{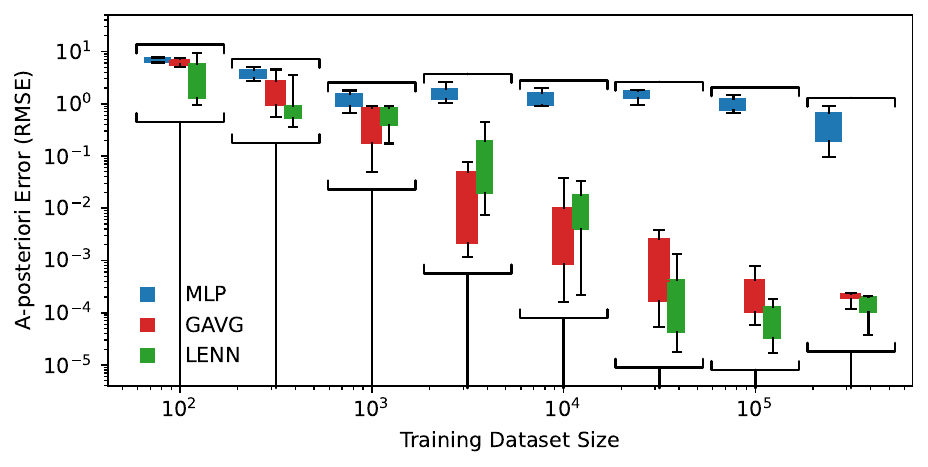}
  \caption{Comparison of the a-posteriori error as a function of the size of the training dataset  for the non-equivariant MLP (blue), MLP with group averaging (GAVG) (red) and LENN (green), for the 2D case. The a-posteriori error is measured as average Relative Mean Square Error (RMSE) on the velocity (Eq.~\ref{eq:aposteriori}). The barplots are computed considering 20 different architectures trained independently. Imposing equivariance in the models enables to reach improved accuracy with smaller training set sizes, showcasing improved parameter efficiency.}
  \label{fig:apo-dset}
\end{figure}

As a second test case, we consider a 3D freely decaying turbulent flow. Here, due to the chaotic nature of turbulence, the networks need to model effectively a much wider set of pre-post collision pairs than in the laminar cases. Thus, this is not only an inherently challenging test case, but it further establishes the wide range of applicability of the learned collisional models.
In order to obtain the initial condition used to evaluate the models, we first run a transient phase using as collision operator the ground truth BGK, with a value of $\tau=0.51$ and stationary force of the form:
\begin{align}\label{eq:bulk-force-3D}
  F_x(x,y,z) = A \sin \left( k y \right), \quad\ \  
  F_y(x,y,z) = A \sin \left( k z \right), \quad\  \ 
  F_z(x,y,z) = A \sin \left( k x \right), 
\end{align}
with $A = 5 \times 10^{-6}$, $ k = 2 \pi / L$, and $x,y,z \in [0, L)$.
The force (Eq.~\ref{eq:bulk-force-3D}) is then turned off, and the different neural network models are evaluated (all starting from the same initial configuration).

The results for this experiment are shown in Fig.~\ref{fig:turbulent-3d}, where we consider the same metrics as the laminar cases (cf. Eq.~\ref{eq:errorrate}, Fig.~\ref{fig:taylor-2d},~\ref{fig:taylor-3d}). We can see that, in the turbulent setting, the non-equivariant MLP fails, with all the simulations diverging in around $100$ timesteps. The two equivariant models, instead, give consistently stable dynamics, with the decay rate oscillating around the ground truth BGK value. Here the fact that the dynamics do not closely follow the ground truth is not surprising: as turbulence is chaotic, any small error by the models propagates in time, giving rise to different solutions.  

Finally, Fig.~\ref{fig:apo-dset} compares the a-posteriori error (cf. Eq.~\ref{eq:aposteriori}) for various models in relation to the size of the training dataset.
We can see that, while both equivariant architectures (GAVG and LENN) show a sharp decay in the error starting from relatively small dataset sizes, the non-equivariant MLP has a much smaller error decay, only manifesting at larger sizes of the datasets (the dataset size used for the experiments above contains $10^5$ datapoints). Equivariant networks thus have a more efficient usage of their capacity as they do not need to ``learn'' lattice symmetries, which reflects in the need of fewer training samples or, for fixed amount of training samples, much higher performance ( the results for a-priori error as a function of the size of the training dataset are reported in the Appendix Sec.~\ref{sec:appendix-res}). %

\section{Conclusion}\label{sec:conclusions}
In this work we have introduced a lattice-equivariant neural network architecture, dubbed LENN, aimed at modeling functions defined on lattice cells, and tailored to preserve
the local symmetries inherent to lattice structures.
Considering the finite symmetry group of the lattice, we constrain the weights of each trainable layer of our LENN to 
ensure equivariance with respect to lattice symmetry. This enables efficient implementations with significant 
reduction of computational costs and memory usage with respect to other methods for enforcing equivariance in a neural network,
such as augmentation of the input data through the symmetry transformation or symmetry-group-averaging of the network.

As a prototype of application, we have considered the problem of learning a surrogate model
for Lattice Boltzmann collision operators, employing LENNs to preserve the symmetries of n-cubic lattices in both 2D and 3D.
We have tested the accuracy of models based on LENNs simulating decaying fluid flows in both laminar and turbulent
regimes, showing systematic superior performances over plain, non-equivariant, multilayer perceptron architectures, regardless
the size of the training dataset. For a fair assessment, we compared LENNs with plain MLPs having the same number of free parameters. MLPs need to allocate part of their capacity to learn the effect of each symmetry group action - possibly memorizing them independently. Contrarily, being the actions of the symmetry group architecturally embedded, our LENNs, and similarly group-averaged networks, can allocate their capacity on other non-linearities of the function modeled. 
Most importantly, we showed that LENNs allow to reduce by up to one order of magnitude (in 3D) the computational costs
in comparison of similarly-sized group-averaged neural networks, while matching the same level of accuracy. This opens to
practical utilization of machine learning-augmented Lattice Boltzmann in real-world simulations.

Finally, we remark that our approach naturally extends to potentially any lattice structure other than the cubic 
lattice (hexagon, honeycomb, icosahedron, etc.), with potential applications in fields inherently connected to the
crystal structure of matter such as solid state physics~\cite{bedolla-jop-2020} and chemistry~\cite{keith-cr-2021},
and for enhancing with machine learning other type of stencil-based computations~\cite{boyda-arxiv-2022}, and
even in higher dimensions~\cite{gabbana-pr-2020}.

\section*{Acknowledgments}
A.G. gratefully acknowledge the support of the U.S. Department of Energy through the LANL/LDRD Program under project number 20240740PRD1 and the Center for Non-Linear Studies for this work. This work was partially funded by the Dutch Research Council (NWO) through the UNRAVEL project (with Project No. OCENW.GROOT.2019.044).

\section*{Appendix}

\subsection{Example of a group-averaged network for a 2D lattice}\label{sec:appendix-ga}
\begin{figure}[H]
  \centering
  \begin{overpic}[width =.8\textwidth]{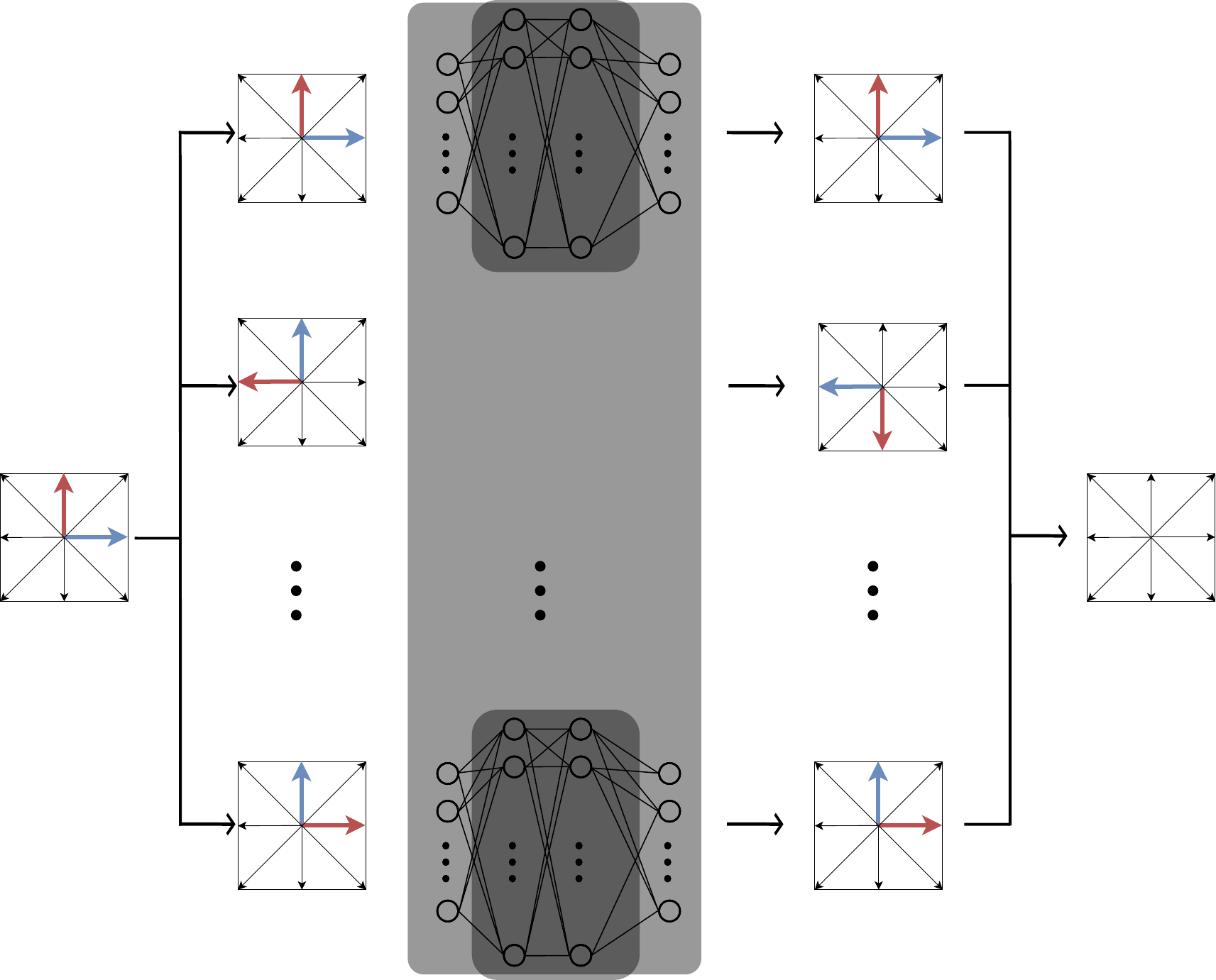}
    \put( 4, 27){\color{black} \large $\mathbf{f}^{\rm pre }$ }
    \put(93, 27){\color{black} \large $\mathbf{\tilde{f}}^{\rm post}$ }
    \put(24, 61){\color{black}        $\mathbf{I}              ~\mathbf{f}^{\rm pre } $  }
    \put(23, 41){\color{black}        $\mathbf{R}              ~\mathbf{f}^{\rm pre } $  }
    \put(22,  4){\color{black}        $\mathbf{R}^3 \mathbf{S} ~\mathbf{f}^{\rm pre } $  }
    \put(70, 61){\color{black}        $\mathbf{I}              ~\mathbf{f}^{\rm post } $  }
    \put(69, 41){\color{black}        $\mathbf{R}^{-1}         ~\mathbf{f}^{\rm post } $  }
    \put(67,  4){\color{black}        $(\mathbf{R}^3 \mathbf{S})^{-1} ~\mathbf{f}^{\rm post } $  }
    \put(40.5, 50  ){\rotatebox{-30}{\color{black} \Large  weight }}
    \put(40.5, 45  ){\rotatebox{-30}{\color{black} \Large  sharing }}
    \put(43.5, 24){\color{black}  \Large   $\Omega^{\rm NN}$}
    \put(78, 38){\color{black} \Large  $\frac{1}{|D_4|} \sum $  }

  \end{overpic} 
  \caption{Sketch of a neural network architecture implementing the group averaging method (Eq.~\ref{eq:groupavg}). 
           The core network, $\Omega^{NN}$ (gray box in the middle), is evaluated $8$ times on 
           rotated/shifted versions of the input. The inverse transformation is applied to
           the $8$ outputs which are then averaged in order to produce the final prediction. 
          }\label{fig:network}
\end{figure}
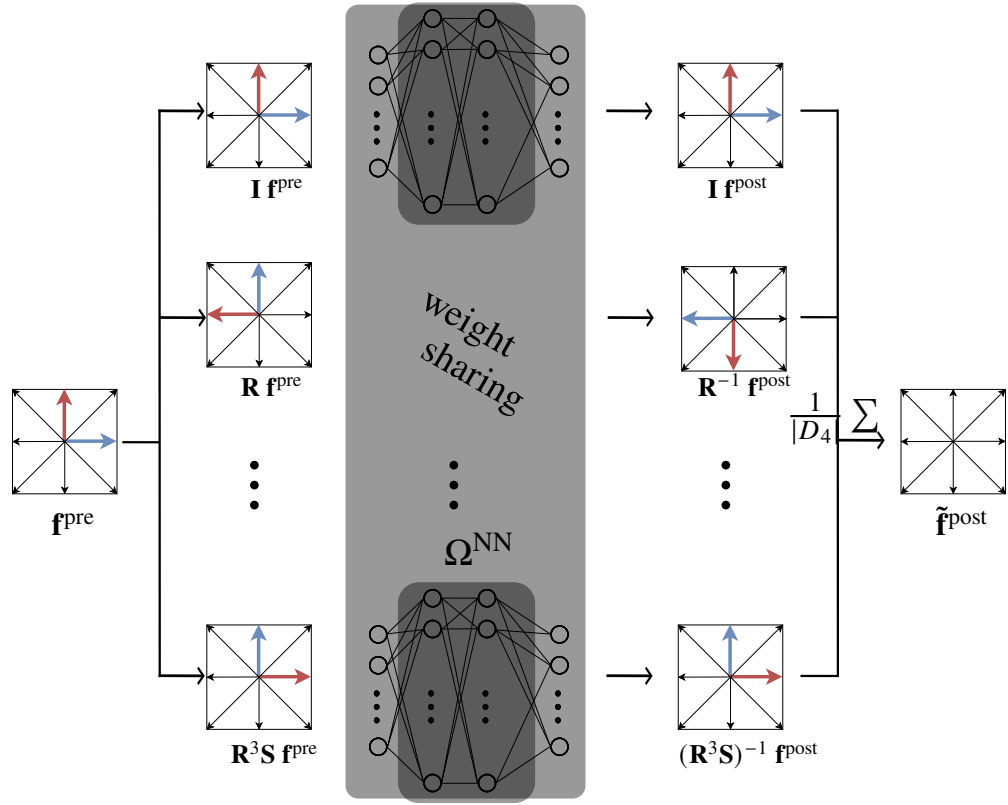

We report here an example of the application of the group averaging approach on a neural network architecture, outlined in
 Sec.~\ref{sec:ga}, considering the 2D case and the D2Q9 stencil (Fig.~\ref{fig:stencils}(a) ), 
and the dihedral symmetry group $D_4$.
The sketch in Fig.~\ref{fig:network} shows the network topology and pipeline description. From left to right, first the 8 different transformations of $D_4$ are applied to the input data (corresponding to the pre-collisional state $\textbf{f}^{\rm pre}$, which is then used to evaluate the core network $\Omega^{\rm NN}$. The 8 different outputs are then transformed with respect to the inverse operation used at the input. Finally, the outputs are averaged to produce the final prediction for the (equivariant) post-collisional state $\tilde{\mathbf{f}}^{\rm post}$.

\newpage
\subsection{Taylor Green vortex: further analysis and 3D case}\label{sec:appendix-res}
\begin{figure}[H]
  \centering
  \includegraphics[width=.8\linewidth]{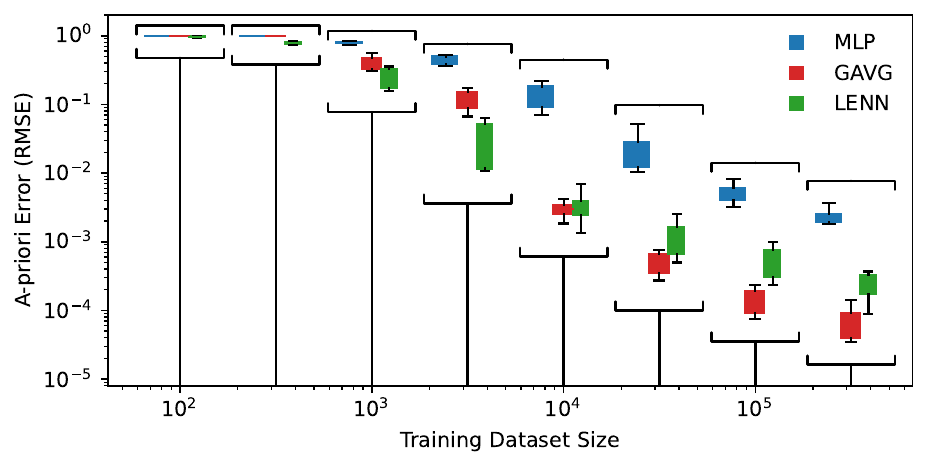}   
  \caption{Comparison of the a-priori error as a function of the size of the training dataset  for the non-equivariant MLP (blue), MLP with group averaging (GAVG) (red) and LENN (green), for the 2D case. The a-priori error is measured as average Relative Mean Square Error (RMSE) on the post-collision populations on the test set (Eq.~\ref{eq:apriori}). The barplots are computed considering 20 different architectures trained independently. Imposing equivariance in the models enables to reach improved accuracy with smaller training set sizes, showcasing improved parameter efficiency and similarly to what observed for the a-posteriori error (cf. Fig.~\ref{fig:apo-dset})}
  \label{fig:apri-dset}
\end{figure}

In the main text, we have evaluated the accuracy in the simulation of the 2D Taylor-Green vortex as a function
of the training dataset size (cf. Fig.~\ref{fig:apri-dset}). For completeness, we report here in Fig.~\ref{fig:apo-dset}
a similar analysis considering this time the a-priori metric defined in Sec.~\ref{sec:res-apo}.
We observe similar trends as those observed in Fig.~\ref{fig:apri-dset} for GAVG and LENN.
At variance with the a-posteriori analysis we see a significant improvement in accuracy for the MLP for larger 
sizes of the training datasets. 

\begin{figure}[H]
 \begin{subfigure}[b]{\linewidth}
  \centering
  \includegraphics[width=\linewidth]{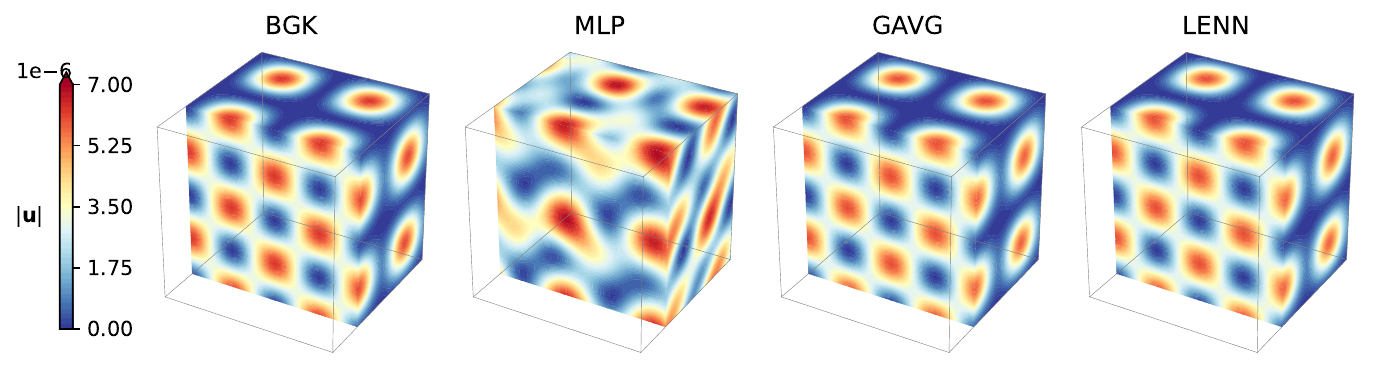}
  \caption{} 
  \vspace{0.2cm}
  \end{subfigure}
  \begin{overpic}[width=\columnwidth]{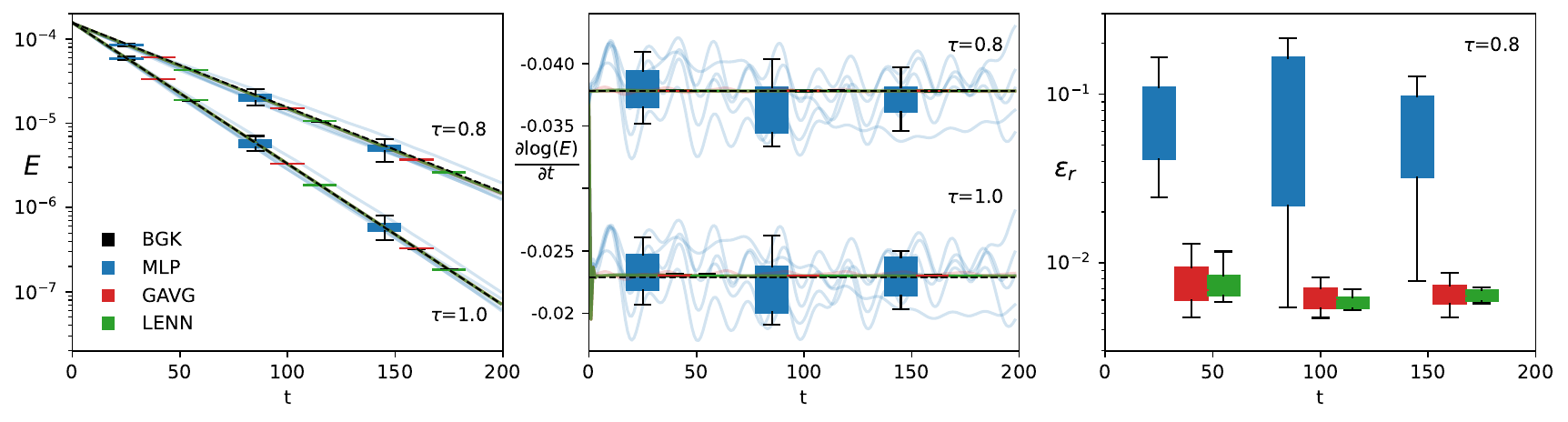}
    \put(15,  -1){\textbf{ (b) }}
    \put(49,  -1){\textbf{ (c) }}
    \put(83,  -1){\textbf{ (d) }}
  \end{overpic}  
  \caption{Results for the 3D Taylor-Green Vortex Decay test case for two different relaxation times $\tau=0.8$, $\tau=1$. 
           Panel (a) shows a qualitative comparison of the velocity magnitude $|\mathbf{u}|$ 
           for the non-equivariant MLP, MLP with group averaging (GAVG) and LENN.
           Panels (b)-(d) illustrate a quantitative comparison for the total energy decay. 
           In panel (b), we show the total energy over time in a logarithmic scale. 
           In panel (c), we show the logarithmic derivative of velocity, representing the local exponential decay rate. 
           In panel (d), we show the relative absolute difference between the predicted local decay rate and 
           the ground truth decay rate, $\epsilon_r$ (Eq.~\ref{eq:errorrate}), for the different models. 
           The barplots are computed considering 20 different architectures trained independently, 
           while for (a) and (b), the individual lines correspond to the evolution for 5 such architectures. 
           The results confirm the method's applicability in the 3D setting, further supporting the considerations made in the 2D setting (cf. Fig.~\ref{fig:taylor-2d}).}
  \label{fig:taylor-3d}
\end{figure}

\noindent For completeness, we also report results for the 3D case, using the following initial condition:
\begin{align}
  & u_x(x,y,z; t_0) = \phantom{-} A \cos\left( k x \right) \sin\left( k y \right) \sin\left( k z \right) \nonumber \\
  & u_y(x,y,z; t_0) = - B \sin\left( k x \right) \cos\left( k y \right) \sin\left( k z \right), \\
  & u_z(x,y,z; t_0) = - B \sin\left( k x \right) \sin\left( k y \right) \cos\left( k z \right) \nonumber
\end{align}
with $A = 0.05$, $B = 0.025$, $k = 2 \pi / L$ and $x,y,z \in [0, L)$.
Once again, we evolve the dynamics using the different neural network models and the ground truth BGK collision. 
The results are reported in Fig.~\ref{fig:taylor-3d}, which are consistent with those reported in the main text for the 2D case,
and further demonstrating the applicability of the proposed model to the 3-dimensional setting.

\newpage
\subsection{Velocity vectors and lattice weights for the D2Q9 and D3Q19 Stencils}\label{sec:appendix-stenc}
The definition of the layers of the LENN architecture reported in the main text depends on the specific labelling 
of the discrete velocities forming the LBM stencil. Therefore, in Table.~\ref{stencil-def-d2q9-d3q19} we report 
the velocity vectors and the lattice weights for the D2Q9 and the D3Q19 models.
\begin{table}[!htb]
    \centering
    \begin{minipage}[t]{0.5\linewidth}
    \setlength\extrarowheight{-5pt}
    \begin{tabular}{ccc}
      Index & $\mathbf{c}_i$ & $w_i$ \\ 
                \hline \\
                $0 $ &  $( 0,  0)$ & $4/9$ \\
                $1 $ &  $( 1,  0)$ & $1/9$ \\
                $2 $ &  $( 0,  1)$ & $1/9$ \\
                $3 $ &  $(-1,  0)$ & $1/9$ \\
                $4 $ &  $( 0, -1)$ & $1/9$ \\
                $5 $ &  $( 1,  1)$ & $1/36$ \\
                $6 $ &  $(-1,  1)$ & $1/36$ \\
                $7 $ &  $(-1, -1)$ & $1/36$ \\
                $8 $ &  $( 1, -1)$ & $1/36$ \\
                     &  \phantom{0}&        \\
                \hline \\
    \end{tabular}
    \end{minipage}
    \hspace{-10em}
    \begin{minipage}[t]{0.5\linewidth}
    \setlength\extrarowheight{-5pt}
    \begin{tabular}{ccc||ccc}
      Index & $\mathbf{c}_i$ & $w_i$ & Index & $\mathbf{c}_i$ & $w_i$ \\ 
        \hline \\
                $0 $ &  $(0, 0, 0)$   & $1/3$  & $10$ &  $(-1, 1, 0)$  & $1/36$ \\
                $1 $ &  $(-1, 0, 0)$  & $1/18$ & $11$ &  $(0, -1, -1)$ & $1/36$ \\
                $2 $ &  $(0, -1, 0)$  & $1/18$ & $12$ &  $(0, -1, 1)$  & $1/36$ \\
                $3 $ &  $(0, 0, -1)$  & $1/18$ & $13$ &  $(0, 1, -1)$  & $1/36$ \\
                $4 $ &  $(0, 0, 1)$   & $1/18$ & $14$ &  $(0, 1, 1)$   & $1/36$ \\
                $5 $ &  $(0, 1, 0)$   & $1/18$ & $15$ &  $(1, -1, 0)$  & $1/36$ \\
                $6 $ &  $(1, 0, 0)$   & $1/18$ & $16$ &  $(1, 0, -1)$  & $1/36$ \\
                $7 $ &  $(-1, -1, 0)$ & $1/36$ & $17$ &  $(1, 0, 1)$   & $1/36$ \\
                $8 $ &  $(-1, 0, -1)$ & $1/36$ & $18$ &  $(1, 1, 0)$   & $1/36$ \\
                $9 $ &  $(-1, 0, 1)$  & $1/36$ &      &                &        \\
        \hline \\
    \end{tabular}
    \end{minipage}
    \caption{Discrete velocity vectors $\mathbf{c}_i$ and corresponding weights $w_i$ for the D2Q9 stencil (left) and the
             D3Q19 stencil (right).}\label{stencil-def-d2q9-d3q19}
\end{table}

\newpage
\subsection{Examples of stencils and their corresponding Equivariant layers}

In this section, we provide the parametrization of the weight matrix $\mathbf{A}$ and of the  bias vector $\mathbf{b}$ satisfying the constraints in Eq.~\ref{eq:equivariantA-constraint} for a few common stencils in 2D and in 3D, respectively: D2Q13, D2Q17, D3Q15, D3Q27.

\subsection*{D2Q13}

\begin{figure}[H]
\centering
\begin{tikzpicture}
\node (centertop) {};
\node (stencil) [left=0.5cm of centertop] {    
\setlength\extrarowheight{-5pt}
\begin{tabular}{ccc}
      Index & $\mathbf{c}_i$ & $w_i$ \\ 
        \hline 
                $0 $ &  $( 0,  0)$ & $3/ 8$ \\
                $1 $ &  $( 1,  0)$ & $1/12$ \\
                $2 $ &  $( 0,  1)$ & $1/12$ \\
                $3 $ &  $(-1,  0)$ & $1/12$ \\
                $4 $ &  $( 0, -1)$ & $1/12$ \\
                $5 $ &  $( 1,  1)$ & $1/16$ \\
                $6 $ &  $(-1,  1)$ & $1/16$ \\
                $7 $ &  $(-1, -1)$ & $1/16$ \\
                $8 $ &  $( 1, -1)$ & $1/16$ \\
                $9 $ &  $( 2,  0)$ & $1/96$ \\
                $10$ &  $( 0,  2)$ & $1/96$ \\
                $11$ &  $(-2,  0)$ & $1/96$ \\
                $12$ &  $( 0, -2)$ & $1/96$ \\
        \hline \\
    \end{tabular}
};
\node (image) [right=1cm of centertop] {
\includegraphics[width=0.45\linewidth]{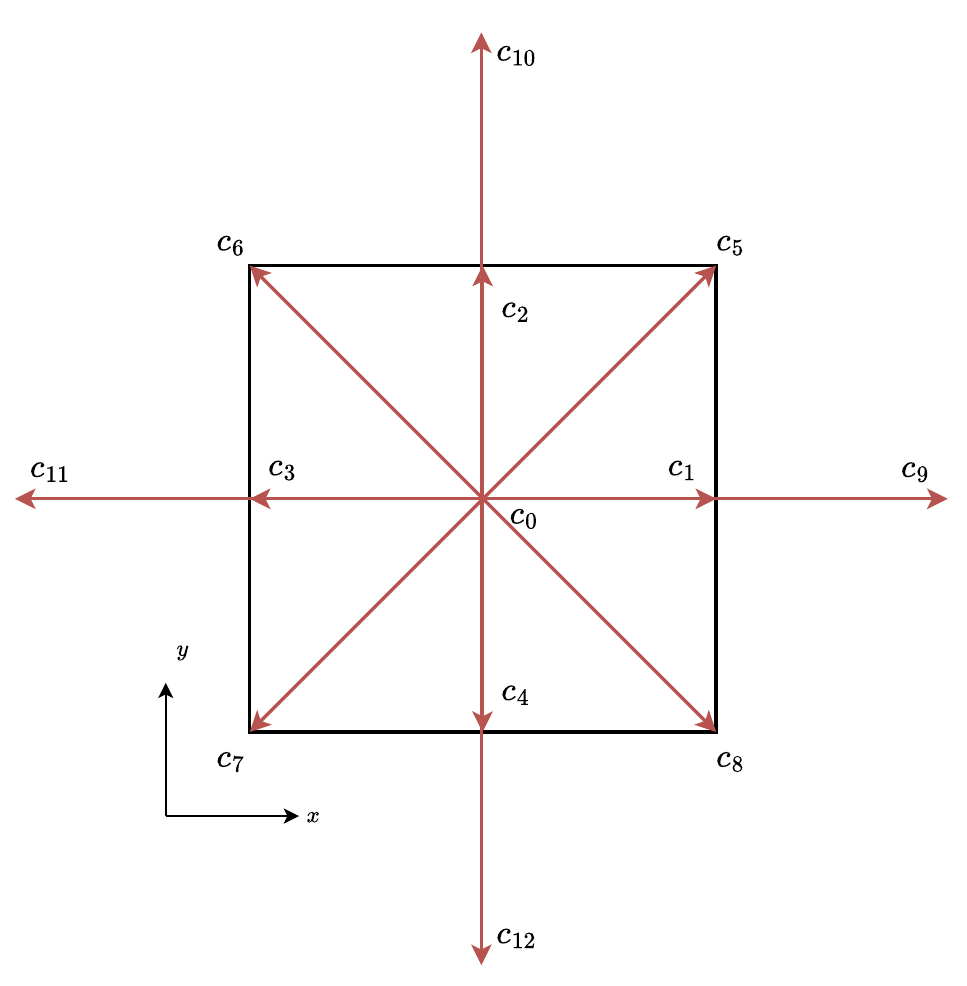}
};
\node (matrix) [below right= 3.95cm and -3cm of  centertop] {
\renewcommand{\arraystretch}{0.65}
\setlength{\arraycolsep}{3pt}
\setcounter{MaxMatrixCols}{20}
$
  \textbf{A}= \left(
  \begin{matrix}
    a & b & b & b & b & d & d & d & d & f & f & f & f \\
     c & h & i & j & i & m & n & n & m & w & x & y & x \\
     c & i & h & i & j & m & m & n & n & x & w & x & y \\
     c & j & i & h & i & n & m & m & n & y & x & w & x \\
     c & i & j & i & h & n & n & m & m & x & y & x & w \\
     e & l & l & k & k & p & q & o & q & z & z & \alpha  & \alpha  \\
     e & k & l & l & k & q & p & q & o & \alpha  & z & z & \alpha  \\
     e & k & k & l & l & o & q & p & q & \alpha  & \alpha  & z & z \\
     e & l & k & k & l & q & o & q & p & z & \alpha  & \alpha  & z \\
     g & r & s & t & s & u & v & v & u & \beta  & \gamma  & \delta  & \gamma  \\
     g & s & r & s & t & u & u & v & v & \gamma  & \beta  & \gamma  & \delta  \\
     g & t & s & r & s & v & u & u & v & \delta  & \gamma  & \beta  & \gamma  \\
     g & s & t & s & r & v & v & u & u & \gamma  & \delta  & \gamma  & \beta  \\
  \end{matrix} \right) \:\:\:\:\:\:\:\:\:\:\:\:
  \textbf{b} = \left( 
  \begin{matrix}
   \epsilon \\
    \zeta \\
    \zeta \\
    \zeta \\
    \zeta \\
    \eta \\
    \eta \\
    \eta \\
    \eta \\
    \theta \\
    \theta \\
    \theta \\
    \theta \\
  \end{matrix} \right).
  $
};    
\draw [] ($(stencil.north west)+(-0.3,0.3)$)  rectangle ($(image.south east)+(0.3,-0.15)$);
\coordinate (SE) at (image.south east);
\coordinate (NW) at (stencil.north west);
\coordinate (MSE) at (matrix.south east);
\path let \p1 = (SE),\p2 = (NW) in coordinate (SW) at (\x2,\y1);
\path let \p1 = (SE),\p2 = (MSE) in coordinate (ASE) at (\x1,\y2);
\draw [] ($(SW)+(-0.3,-0.15)$)  rectangle ($(ASE)+(0.3,-0.15)$);
\end{tikzpicture}
\end{figure}

\subsection*{D2Q17}

\begin{figure}[H]
\centering
\begin{tikzpicture}
\node (centertop) {};
\node (stencil) [left=0.5cm of centertop] {    
\setlength\extrarowheight{-5pt}
    \begin{tabular}{ccc}
      Index & $\mathbf{c}_i$ & $w_i$ \\ \hline
                $0 $ &  $(0, 0)$ & $ (575 + 193\sqrt{193})/8100$ \\  
                $1 $ &  $(1, 0)$ & $(3355 - 91\sqrt{193})/18000$ \\
                $2 $ &  $(0, 1)$ & $(3355 - 91\sqrt{193})/18000$ \\
                $3 $ &  $(-1, 0)$ & $(3355 - 91\sqrt{193})/18000$ \\
                $4 $ &  $(0, -1)$ & $(3355 - 91\sqrt{193})/18000$ \\
                $5 $ &  $(1, 1)$ & $(655 + 17\sqrt{193})/27000$ \\
                $6 $ &  $(-1, 1)$ & $(655 + 17\sqrt{193})/27000$ \\
                $7 $ &  $(-1, -1)$ & $(655 + 17\sqrt{193})/27000$ \\
                $8 $ &  $(1, -1)$ & $(655 + 17\sqrt{193})/27000$ \\
                $9 $ &  $(3, 0)$ & $(1445 - 101\sqrt{193})/162000$ \\
                $10$ &  $(0, 3)$ & $(1445 - 101\sqrt{193})/162000$ \\
                $11$ &  $(-3, 0)$ & $(1445 - 101\sqrt{193})/162000$ \\
                $12$ &  $(0, -3)$ & $(1445 - 101\sqrt{193})/162000$ \\
                $13$ &  $(2, 2)$ & $(685 - 49\sqrt{193})/54000$ \\
                $14$ &  $(-2, 2)$ & $(685 - 49\sqrt{193})/54000$ \\
                $15$ &  $(-2, -2)$ & $(685 - 49\sqrt{193})/54000$ \\
                $16$ &  $(2, -2)$ & $(685 - 49\sqrt{193})/54000$ \\
                \hline \\
    \end{tabular}
};
\node (image) [right=1cm of centertop] {
    \includegraphics[width=0.45\linewidth]{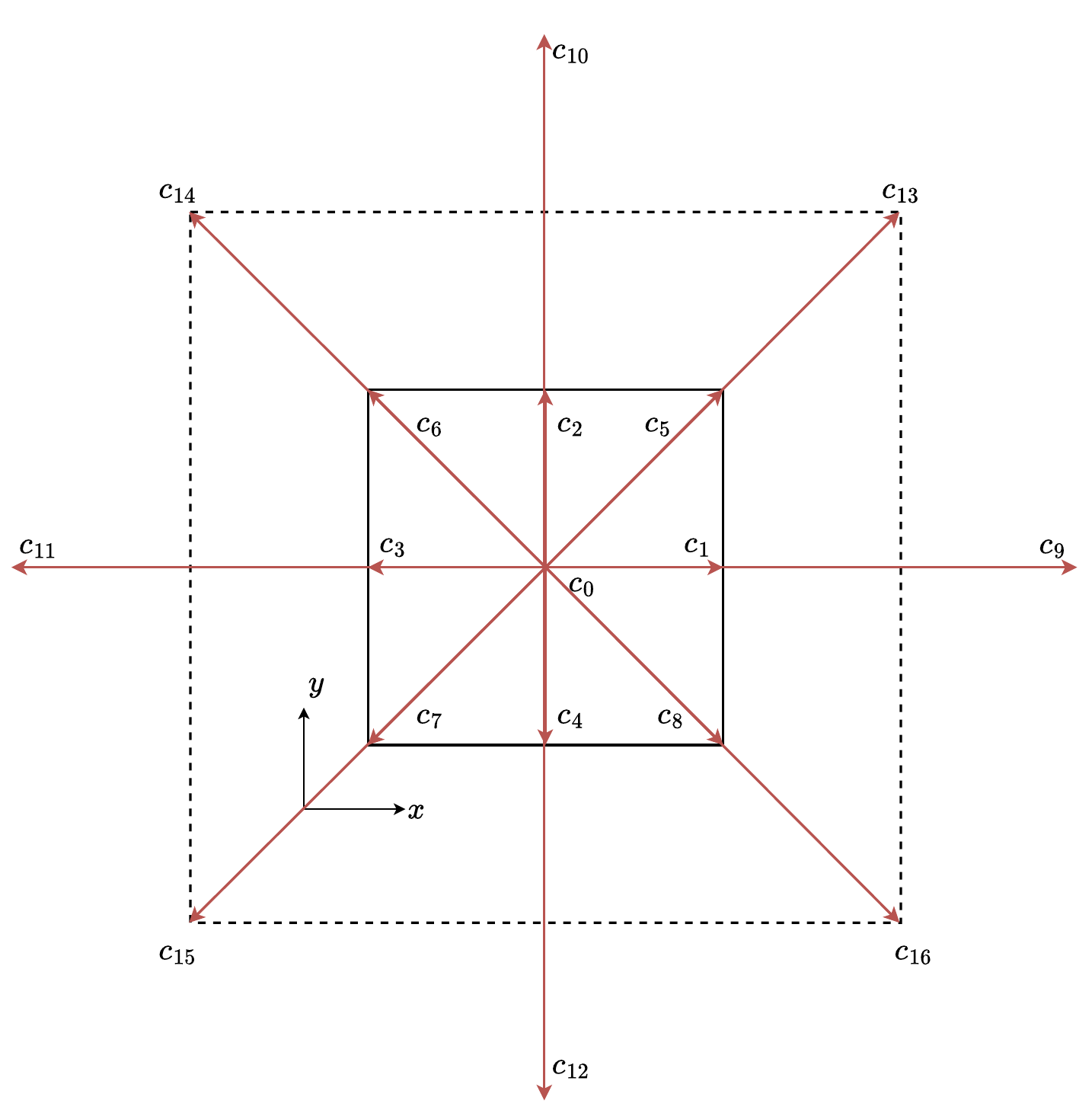}
};
\node (matrix) [below right= 5cm and -6cm of  centertop] {
\renewcommand{\arraystretch}{0.65}
\setlength{\arraycolsep}{3pt}
\setcounter{MaxMatrixCols}{50}
$
  \textbf{A}= \left(
  \begin{matrix}
    a & b & b & b & b & d & d & d & d & f & f & f & f & h & h & h & h \\
     c & j & k & l & k & m & n & n & m & q & r & s & r & z & \alpha  & \alpha  & z \\
     c & k & j & k & l & m & m & n & n & r & q & r & s & z & z & \alpha  & \alpha  \\
     c & l & k & j & k & n & m & m & n & s & r & q & r & \alpha  & z & z & \alpha  \\
     c & k & l & k & j & n & n & m & m & r & s & r & q & \alpha  & \alpha  & z & z \\
     e & o & o & p & p & t & u & v & u & \beta  & \beta  & \gamma  & \gamma  & \theta  & \kappa  & \lambda  & \kappa  \\
     e & p & o & o & p & u & t & u & v & \gamma  & \beta  & \beta  & \gamma  & \kappa  & \theta  & \kappa  & \lambda  \\
     e & p & p & o & o & v & u & t & u & \gamma  & \gamma  & \beta  & \beta  & \lambda  & \kappa  & \theta  & \kappa  \\
     e & o & p & p & o & u & v & u & t & \beta  & \gamma  & \gamma  & \beta  & \kappa  & \lambda  & \kappa  & \theta  \\
     g & w & x & y & x & \delta  & \epsilon  & \epsilon  & \delta  & \mu  & \xi  & \pi  & \xi  & \phi  & \chi  & \chi  & \phi  \\
     g & x & w & x & y & \delta  & \delta  & \epsilon  & \epsilon  & \xi  & \mu  & \xi  & \pi  & \phi  & \phi  & \chi  & \chi  \\
     g & y & x & w & x & \epsilon  & \delta  & \delta  & \epsilon  & \pi  & \xi  & \mu  & \xi  & \chi  & \phi  & \phi  & \chi  \\
     g & x & y & x & w & \epsilon  & \epsilon  & \delta  & \delta  & \xi  & \pi  & \xi  & \mu  & \chi  & \chi  & \phi  & \phi  \\
     i & \zeta  & \zeta  & \eta  & \eta  & \rho  & \sigma  & \upsilon  & \sigma  & \psi  & \psi  & \omega  & \omega  & \digamma  & \varsigma & \varepsilon & \varsigma \\
     i & \eta  & \zeta  & \zeta  & \eta  & \sigma  & \rho  & \sigma  & \upsilon  & \omega  & \psi  & \psi  & \omega  & \varsigma & \digamma  & \varsigma & \varepsilon \\
     i & \eta  & \eta  & \zeta  & \zeta  & \upsilon  & \sigma  & \rho  & \sigma  & \omega  & \omega  & \psi  & \psi  & \varepsilon & \varsigma & \digamma  & \varsigma \\
     i & \zeta  & \eta  & \eta  & \zeta  & \sigma  & \upsilon  & \sigma  & \rho  & \psi  & \omega  & \omega  & \psi  & \varsigma & \varepsilon & \varsigma & \digamma  \\
  \end{matrix} \right) \:\:\:\:\:\:\:\:\:\:\:\:
  \textbf{b} = \left( 
  \begin{matrix}
    \vartheta \\
    \varpi \\
    \varpi \\
    \varpi \\
    \varpi \\
    \varrho \\
    \varrho \\
    \varrho \\
    \varrho \\
    \varphi \\
    \varphi \\
    \varphi \\
    \varphi \\
    \varphi \\
    \varkappa \\
    \varkappa \\
    \varkappa \\
    \varkappa \\
    \varkappa \\
  \end{matrix} \right).
  $
};    
\draw [] ($(stencil.north west)+(-0.3,0.3)$)  rectangle ($(image.south east)+(0.3,-1.0)$);
\coordinate (SE) at (image.south east);
\coordinate (NW) at (stencil.north west);
\coordinate (MSE) at (matrix.south east);
\path let \p1 = (SE),\p2 = (NW) in coordinate (SW) at (\x2,\y1);
\path let \p1 = (SE),\p2 = (MSE) in coordinate (ASE) at (\x1,\y2);
\draw [] ($(SW)+(-0.3,-1.0)$)  rectangle ($(ASE)+(0.3,-1.0)$);
\end{tikzpicture}
\end{figure}

\subsection*{D3Q15}

\begin{figure}[H]
\centering
\begin{tikzpicture}
\node (centertop) {};
\node (stencil) [left=0.5cm of centertop] {    
\setlength\extrarowheight{-5pt}
    \begin{tabular}{ccc}
      Index & $\mathbf{c}_i$ & $w_i$ \\ \hline
                $0 $ &  $( 0,  0,  0)$ & $2 /  9$ \\
                $1 $ &  $(-1,  0,  0)$ & $1 /  9$ \\
                $2 $ &  $( 0, -1,  0)$ & $1 /  9$ \\
                $3 $ &  $( 0,  0, -1)$ & $1 /  9$ \\
                $4 $ &  $( 0,  0,  1)$ & $1 /  9$ \\
                $5 $ &  $( 0,  1,  0)$ & $1 /  9$ \\
                $6 $ &  $( 1,  0,  0)$ & $1 /  9$ \\
                $7 $ &  $(-1, -1, -1)$ & $1 / 72$ \\
                $8 $ &  $(-1, -1,  1)$ & $1 / 72$ \\
                $9 $ &  $(-1,  1, -1)$ & $1 / 72$ \\
                $10$ &  $(-1,  1,  1)$ & $1 / 72$ \\
                $11$ &  $( 1, -1, -1)$ & $1 / 72$ \\
                $12$ &  $( 1, -1,  1)$ & $1 / 72$ \\
                $13$ &  $( 1,  1, -1)$ & $1 / 72$ \\
                $14$ &  $( 1,  1,  1)$ & $1 / 72$ \\
                \hline \\
    \end{tabular}
};
\node (image) [right=1cm of centertop] {
    \includegraphics[width=0.45\linewidth]{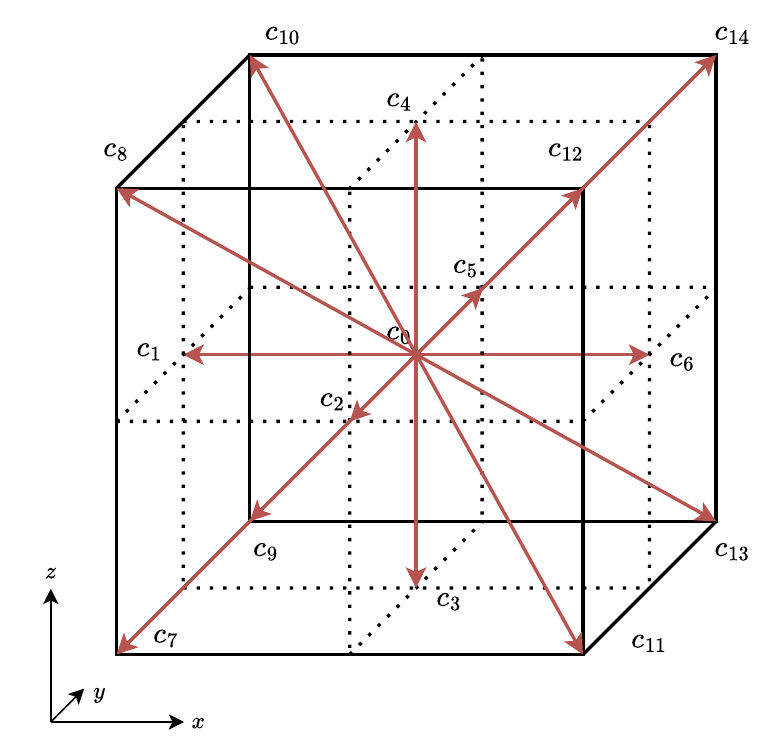}
};
\node (matrix) [below right= 5cm and -4cm of  centertop] {
\renewcommand{\arraystretch}{0.65}
\setlength{\arraycolsep}{3pt}
\setcounter{MaxMatrixCols}{50}
$
  \textbf{A}= \left(
  \begin{matrix}
      a & b & b & b & b & b & b & c & c & c & c & c & c & c & c \\
      d & f & g & g & g & g & h & i & i & i & i & j & j & j & j \\
      d & g & f & g & g & h & g & i & i & j & j & i & i & j & j \\
      d & g & g & f & h & g & g & i & j & i & j & i & j & i & j \\
      d & g & g & h & f & g & g & j & i & j & i & j & i & j & i \\
      d & g & h & g & g & f & g & j & j & i & i & j & j & i & i \\
      d & h & g & g & g & g & f & j & j & j & j & i & i & i & i \\
      e & l & l & l & k & k & k & m & p & p & o & p & o & o & n \\
      e & l & l & k & l & k & k & p & m & o & p & o & p & n & o \\
      e & l & k & l & k & l & k & p & o & m & p & o & n & p & o \\
      e & l & k & k & l & l & k & o & p & p & m & n & o & o & p \\
      e & k & l & l & k & k & l & p & o & o & n & m & p & p & o \\
      e & k & l & k & l & k & l & o & p & n & o & p & m & o & p \\
      e & k & k & l & k & l & l & o & n & p & o & p & o & m & p \\
      e & k & k & k & l & l & l & n & o & o & p & o & p & p & m \\ 
    \end{matrix} \right) \:\:\:\:\:\:\:\:\:\:\:\:
  \textbf{b} = \left( 
  \begin{matrix}
    q \\
    r \\
    r \\
    r \\
    r \\
    r \\
    r \\
    s \\
    s \\
    s \\
    s \\
    s \\
    s \\
    s \\
    s \\
  \end{matrix} \right).
  $
};    
\draw [] ($(stencil.north west)+(-0.3,0.3)$)  rectangle ($(image.south east)+(0.3,-1.0)$);
\coordinate (SE) at (image.south east);
\coordinate (NW) at (stencil.north west);
\coordinate (MSE) at (matrix.south east);
\path let \p1 = (SE),\p2 = (NW) in coordinate (SW) at (\x2,\y1);
\path let \p1 = (SE),\p2 = (MSE) in coordinate (ASE) at (\x1,\y2);
\draw [] ($(SW)+(-0.3,-1.0)$)  rectangle ($(ASE)+(0.3,-1.0)$);
\end{tikzpicture}
\end{figure}

\subsection*{D3Q27}
\begin{figure}[H]
\centering
\begin{tikzpicture}
\node (centertop) {};
\node (stencil) [left=-1.5cm of centertop] {    
\setlength\extrarowheight{-5pt}
    \begin{tabular}{ccc||ccc}
      Index & $\mathbf{c}_i$ & $w_i$ & Index &$\mathbf{c}_i$ & $w_i$ \\ \hline
                $0 $ &  $( 0,  0,  0)$ & $8/27 $ & $14$ &  $( 0,  1,  1)$ & $1/54 $ \\  
                $1 $ &  $(-1,  0,  0)$ & $2/27 $ & $15$ &  $( 1, -1,  0)$ & $1/54 $ \\
                $2 $ &  $( 0, -1,  0)$ & $2/27 $ & $16$ &  $( 1,  0, -1)$ & $1/54 $ \\
                $3 $ &  $( 0,  0, -1)$ & $2/27 $ & $17$ &  $( 1,  0,  1)$ & $1/54 $ \\
                $4 $ &  $( 0,  0,  1)$ & $2/27 $ & $18$ &  $( 1,  1,  0)$ & $1/54 $ \\
                $5 $ &  $( 0,  1,  0)$ & $2/27 $ & $19$ &  $(-1, -1, -1)$ & $1/216$ \\
                $6 $ &  $( 1,  0,  0)$ & $2/27 $ & $20$ &  $(-1, -1,  1)$ & $1/216$ \\
                $7 $ &  $(-1, -1,  0)$ & $1/54 $ & $21$ &  $(-1,  1, -1)$ & $1/216$ \\
                $8 $ &  $(-1,  0, -1)$ & $1/54 $ & $22$ &  $(-1,  1,  1)$ & $1/216$ \\
                $9 $ &  $(-1,  0,  1)$ & $1/54 $ & $23$ &  $( 1, -1, -1)$ & $1/216$ \\
                $10$ &  $(-1,  1,  0)$ & $1/54 $ & $24$ &  $( 1, -1,  1)$ & $1/216$ \\
                $11$ &  $( 0, -1, -1)$ & $1/54 $ & $25$ &  $( 1,  1, -1)$ & $1/216$ \\
                $12$ &  $( 0, -1,  1)$ & $1/54 $ & $26$ &  $( 1,  1,  1)$ & $1/216$ \\
                $13$ &  $( 0,  1, -1)$ & $1/54 $ &      &  \phantom{$0$}    &       \\
                \hline \\
    \end{tabular}
};
\node (image) [right=1.5cm of centertop] {
    \includegraphics[width=0.45\linewidth]{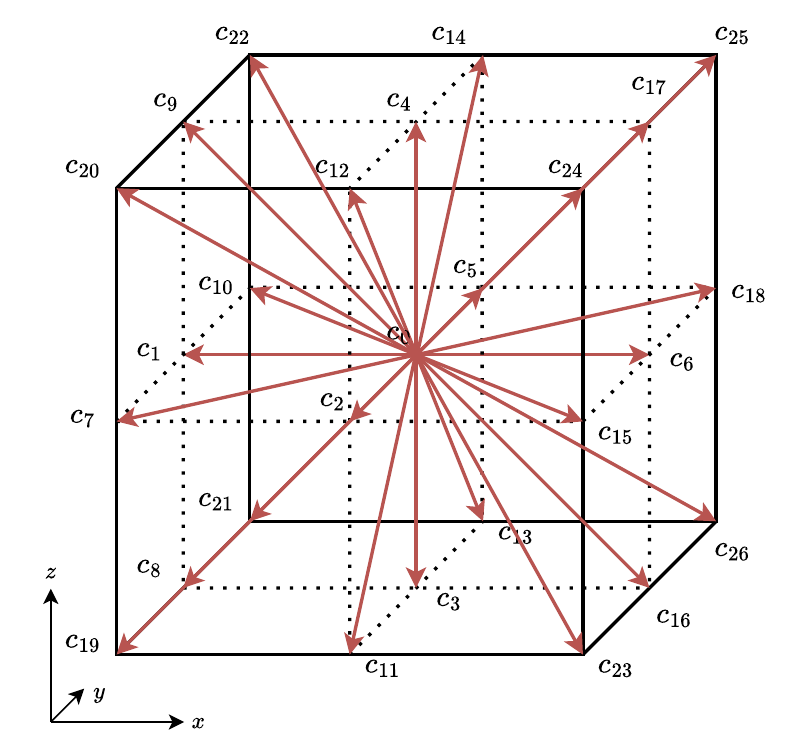}
};
\node (matrix) [below right= 5cm and -6cm of  centertop] {
\renewcommand{\arraystretch}{0.65}
\setlength{\arraycolsep}{3pt}
\setcounter{MaxMatrixCols}{50}
$
  \textbf{A}= \left(
  \begin{smallmatrix}
    a & b & b & b & b & b & b & d & d & d & d & d & d & d & d & d & d & d & d & f & f & f & f & f & f & f & f \\
     c & h & i & i & i & i & j & k & k & k & k & l & l & l & l & m & m & m & m & v & v & v & v & w & w & w & w \\
     c & i & h & i & i & j & i & k & l & l & m & k & k & m & m & k & l & l & m & v & v & w & w & v & v & w & w \\
     c & i & i & h & j & i & i & l & k & m & l & k & m & k & m & l & k & m & l & v & w & v & w & v & w & v & w \\
     c & i & i & j & h & i & i & l & m & k & l & m & k & m & k & l & m & k & l & w & v & w & v & w & v & w & v \\
     c & i & j & i & i & h & i & m & l & l & k & m & m & k & k & m & l & l & k & w & w & v & v & w & w & v & v \\
     c & j & i & i & i & i & h & m & m & m & m & l & l & l & l & k & k & k & k & w & w & w & w & v & v & v & v \\
     e & n & n & o & o & p & p & q & r & r & s & r & r & t & t & s & t & t & u & z & z & \alpha  & \alpha  & \alpha  & \alpha  & \beta  & \beta  \\
     e & n & o & n & p & o & p & r & q & s & r & r & t & r & t & t & s & u & t & z & \alpha  & z & \alpha  & \alpha  & \beta  & \alpha  & \beta  \\
     e & n & o & p & n & o & p & r & s & q & r & t & r & t & r & t & u & s & t & \alpha  & z & \alpha  & z & \beta  & \alpha  & \beta  & \alpha  \\
     e & n & p & o & o & n & p & s & r & r & q & t & t & r & r & u & t & t & s & \alpha  & \alpha  & z & z & \beta  & \beta  & \alpha  & \alpha  \\
     e & o & n & n & p & p & o & r & r & t & t & q & s & s & u & r & r & t & t & z & \alpha  & \alpha  & \beta  & z & \alpha  & \alpha  & \beta  \\
     e & o & n & p & n & p & o & r & t & r & t & s & q & u & s & r & t & r & t & \alpha  & z & \beta  & \alpha  & \alpha  & z & \beta  & \alpha  \\
     e & o & p & n & p & n & o & t & r & t & r & s & u & q & s & t & r & t & r & \alpha  & \beta  & z & \alpha  & \alpha  & \beta  & z & \alpha  \\
     e & o & p & p & n & n & o & t & t & r & r & u & s & s & q & t & t & r & r & \beta  & \alpha  & \alpha  & z & \beta  & \alpha  & \alpha  & z \\
     e & p & n & o & o & p & n & s & t & t & u & r & r & t & t & q & r & r & s & \alpha  & \alpha  & \beta  & \beta  & z & z & \alpha  & \alpha  \\
     e & p & o & n & p & o & n & t & s & u & t & r & t & r & t & r & q & s & r & \alpha  & \beta  & \alpha  & \beta  & z & \alpha  & z & \alpha  \\
     e & p & o & p & n & o & n & t & u & s & t & t & r & t & r & r & s & q & r & \beta  & \alpha  & \beta  & \alpha  & \alpha  & z & \alpha  & z \\
     e & p & p & o & o & n & n & u & t & t & s & t & t & r & r & s & r & r & q & \beta  & \beta  & \alpha  & \alpha  & \alpha  & \alpha  & z & z \\
     g & x & x & x & y & y & y & \gamma  & \gamma  & \delta  & \delta  & \gamma  & \delta  & \delta  & \epsilon  & \delta  & \delta  & \epsilon  & \epsilon  & \zeta  & \eta  & \eta  & \theta  & \eta  & \theta  & \theta  & \kappa  \\
     g & x & x & y & x & y & y & \gamma  & \delta  & \gamma  & \delta  & \delta  & \gamma  & \epsilon  & \delta  & \delta  & \epsilon  & \delta  & \epsilon  & \eta  & \zeta  & \theta  & \eta  & \theta  & \eta  & \kappa  & \theta  \\
     g & x & y & x & y & x & y & \delta  & \gamma  & \delta  & \gamma  & \delta  & \epsilon  & \gamma  & \delta  & \epsilon  & \delta  & \epsilon  & \delta  & \eta  & \theta  & \zeta  & \eta  & \theta  & \kappa  & \eta  & \theta  \\
     g & x & y & y & x & x & y & \delta  & \delta  & \gamma  & \gamma  & \epsilon  & \delta  & \delta  & \gamma  & \epsilon  & \epsilon  & \delta  & \delta  & \theta  & \eta  & \eta  & \zeta  & \kappa  & \theta  & \theta  & \eta  \\
     g & y & x & x & y & y & x & \delta  & \delta  & \epsilon  & \epsilon  & \gamma  & \delta  & \delta  & \epsilon  & \gamma  & \gamma  & \delta  & \delta  & \eta  & \theta  & \theta  & \kappa  & \zeta  & \eta  & \eta  & \theta  \\
     g & y & x & y & x & y & x & \delta  & \epsilon  & \delta  & \epsilon  & \delta  & \gamma  & \epsilon  & \delta  & \gamma  & \delta  & \gamma  & \delta  & \theta  & \eta  & \kappa  & \theta  & \eta  & \zeta  & \theta  & \eta  \\
     g & y & y & x & y & x & x & \epsilon  & \delta  & \epsilon  & \delta  & \delta  & \epsilon  & \gamma  & \delta  & \delta  & \gamma  & \delta  & \gamma  & \theta  & \kappa  & \eta  & \theta  & \eta  & \theta  & \zeta  & \eta  \\
     g & y & y & y & x & x & x & \epsilon  & \epsilon  & \delta  & \delta  & \epsilon  & \delta  & \delta  & \gamma  & \delta  & \delta  & \gamma  & \gamma  & \kappa  & \theta  & \theta  & \eta  & \theta  & \eta  & \eta  & \zeta  \\
  \end{smallmatrix} \right) \:\:\:\:\:\:\:\:\:\:\:\:
  \textbf{b} = \left( 
  \begin{smallmatrix}
    \lambda \\
    \mu \\
    \mu \\
    \mu \\
    \mu \\
    \mu \\
    \mu \\
    \nu \\
    \nu \\
    \nu \\
    \nu \\
    \nu \\
    \nu \\
    \nu \\
    \nu \\
    \nu \\
    \nu \\
    \nu \\
    \nu \\
    \xi \\
    \xi \\
    \xi \\
    \xi \\
    \xi \\
    \xi \\    
    \xi \\
    \xi \\    
  \end{smallmatrix} \right).
  $
};    
\draw [] ($(stencil.north west)+(-0.3,0.3)$)  rectangle ($(image.south east)+(0.3,-1.0)$);
\coordinate (SE) at (image.south east);
\coordinate (NW) at (stencil.north west);
\coordinate (MSE) at (matrix.south east);
\path let \p1 = (SE),\p2 = (NW) in coordinate (SW) at (\x2,\y1);
\path let \p1 = (SE),\p2 = (MSE) in coordinate (ASE) at (\x1,\y2);
\draw [] ($(SW)+(-0.3,-1.0)$)  rectangle ($(ASE)+(0.3,-1.0)$);
\end{tikzpicture}
\end{figure}

\newpage
\bibliography{biblio}

\begin{thebibliography}{37}
\newcommand{\enquote}[1]{``#1''}
\providecommand{\natexlab}[1]{#1}
\providecommand{\url}[1]{\texttt{#1}}
\providecommand{\urlprefix}{URL }
\expandafter\ifx\csname urlstyle\endcsname\relax
  \providecommand{\doi}[1]{\discretionary{}{}{}https://doi.org/#1}\else
  \providecommand{\doi}[1]{\discretionary{}{}{}\urlstyle{rm}\url{https://doi.org/#1}}\fi

\bibitem[{Karniadakis et~al.(2021{\natexlab{a}})Karniadakis, Kevrekidis, Lu,
  Perdikaris, Wang, and Yang}]{osti_1852843}
Karniadakis, G.~E., Kevrekidis, I.~G., Lu, L., Perdikaris, P., Wang, S., and
  Yang, L., \enquote{Physics-informed machine learning,} \emph{Nature Reviews
  Physics}, Vol.~3, No.~6, 2021{\natexlab{a}}.
\newblock \doi{10.1038/s42254-021-00314-5}.

\bibitem[{Carleo et~al.(2019)Carleo, Cirac, Cranmer, Daudet, Schuld, Tishby,
  Vogt-Maranto, and Zdeborov\'a}]{RevModPhys.91.045002}
Carleo, G., Cirac, I., Cranmer, K., Daudet, L., Schuld, M., Tishby, N.,
  Vogt-Maranto, L., and Zdeborov\'a, L., \enquote{Machine learning and the
  physical sciences,} \emph{Rev. Mod. Phys.}, Vol.~91, 2019, p. 045002.
\newblock \doi{10.1103/RevModPhys.91.045002}.

\bibitem[{Karniadakis et~al.(2021{\natexlab{b}})Karniadakis, Kevrekidis, Lu,
  Perdikaris, Wang, and Yang}]{karniadakis-natrev-2021}
Karniadakis, G.~E., Kevrekidis, I.~G., Lu, L., Perdikaris, P., Wang, S., and
  Yang, L., \enquote{Physics-informed machine learning,} \emph{Nature Review
  Physics}, Vol.~3, No.~6, 2021{\natexlab{b}}, pp. 422--440.
\newblock \doi{10.1038/s42254-021-00314-5}.

\bibitem[{Noether(1918)}]{noether-1918}
Noether, E., \enquote{Invariante Variationsprobleme,} \emph{Nachrichten von der
  Gesellschaft der Wissenschaften zu Göttingen, Mathematisch-Physikalische
  Klasse}, Vol. 1918, 1918, pp. 235--257.
\newblock \urlprefix\url{http://eudml.org/doc/59024}.

\bibitem[{Cohen and Welling(2016)}]{cohen-icml-2016}
Cohen, T., and Welling, M., \enquote{Group equivariant convolutional networks,}
  \emph{International conference on machine learning}, PMLR, 2016, pp.
  2990--2999.
\newblock \doi{10.48550/arXiv.1602.07576}.

\bibitem[{Bronstein et~al.(2021)Bronstein, Bruna, Cohen, and
  Veli{\v{c}}kovi{\'c}}]{bronstein-arxiv-2021}
Bronstein, M.~M., Bruna, J., Cohen, T., and Veli{\v{c}}kovi{\'c}, P.,
  \enquote{Geometric deep learning: Grids, groups, graphs, geodesics, and
  gauges,} \emph{arXiv preprint arXiv:2104.13478}, 2021.
\newblock \doi{10.48550/arXiv.2104.13478}.

\bibitem[{Cubuk et~al.(2018)Cubuk, Zoph, Mane, Vasudevan, and
  Le}]{cubuk-arxiv-2018}
Cubuk, E.~D., Zoph, B., Mane, D., Vasudevan, V., and Le, Q.~V.,
  \enquote{Autoaugment: Learning augmentation policies from data,} \emph{arXiv
  preprint arXiv:1805.09501}, 2018.
\newblock \doi{10.48550/arXiv.1805.09501}.

\bibitem[{Lorraine et~al.(2020)Lorraine, Vicol, and
  Duvenaud}]{lorraine-icai-2020}
Lorraine, J., Vicol, P., and Duvenaud, D., \enquote{Optimizing millions of
  hyperparameters by implicit differentiation,} \emph{International conference
  on artificial intelligence and statistics}, PMLR, 2020, pp. 1540--1552.
\newblock \doi{10.48550/arXiv.1911.02590}.

\bibitem[{Laptev et~al.(2016)Laptev, Savinov, Buhmann, and
  Pollefeys}]{laptev-ieee-2016}
Laptev, D., Savinov, N., Buhmann, J.~M., and Pollefeys, M.,
  \enquote{Ti-pooling: transformation-invariant pooling for feature learning in
  convolutional neural networks,} \emph{Proceedings of the IEEE conference on
  computer vision and pattern recognition}, 2016, pp. 289--297.
\newblock \doi{10.48550/arXiv.1604.06318}.

\bibitem[{Fukushima(1980)}]{fukushima-bc-1980}
Fukushima, K., \enquote{Neocognitron: A self-organizing neural network model
  for a mechanism of pattern recognition unaffected by shift in position,}
  \emph{Biological cybernetics}, Vol.~36, No.~4, 1980, pp. 193--202.
\newblock \doi{10.1007/BF00344251}.

\bibitem[{LeCun et~al.(1989)LeCun, Boser, Denker, Henderson, Howard, Hubbard,
  and Jackel}]{lecun-nc-1989}
LeCun, Y., Boser, B., Denker, J.~S., Henderson, D., Howard, R.~E., Hubbard, W.,
  and Jackel, L.~D., \enquote{Backpropagation applied to handwritten zip code
  recognition,} \emph{Neural computation}, Vol.~1, No.~4, 1989, pp. 541--551.
\newblock \doi{10.1162/neco.1989.1.4.541}.

\bibitem[{Krizhevsky et~al.(2012)Krizhevsky, Sutskever, and
  Hinton}]{NIPS2012_c399862d}
Krizhevsky, A., Sutskever, I., and Hinton, G.~E., \enquote{ImageNet
  Classification with Deep Convolutional Neural Networks,} \emph{Advances in
  Neural Information Processing Systems}, Vol.~25, edited by F.~Pereira,
  C.~Burges, L.~Bottou, and K.~Weinberger, Curran Associates, Inc., 2012.
\newblock
  \urlprefix\url{https://proceedings.neurips.cc/paper_files/paper/2012/file/c399862d3b9d6b76c8436e924a68c45b-Paper.pdf}.

\bibitem[{Ravanbakhsh et~al.(2017)Ravanbakhsh, Schneider, and
  Poczos}]{ravanbakhsh-icml-2017}
Ravanbakhsh, S., Schneider, J., and Poczos, B., \enquote{Equivariance through
  parameter-sharing,} \emph{International conference on machine learning},
  PMLR, 2017, pp. 2892--2901.
\newblock \doi{10.48550/arXiv.1702.08389}.

\bibitem[{Kondor and Trivedi(2018)}]{kondor-icml-2018}
Kondor, R., and Trivedi, S., \enquote{On the generalization of equivariance and
  convolution in neural networks to the action of compact groups,}
  \emph{International conference on machine learning}, PMLR, 2018, pp.
  2747--2755.
\newblock \doi{10.48550/arXiv.1802.03690}.

\bibitem[{Weiler et~al.(2021)Weiler, Forr{\'e}, Verlinde, and
  Welling}]{weiler-arxiv-2021}
Weiler, M., Forr{\'e}, P., Verlinde, E., and Welling, M., \enquote{Coordinate
  Independent Convolutional Networks--Isometry and Gauge Equivariant
  Convolutions on Riemannian Manifolds,} \emph{arXiv preprint
  arXiv:2106.06020}, 2021.
\newblock \doi{10.48550/arXiv.2106.06020}.

\bibitem[{Finzi et~al.(2021)Finzi, Welling, and Wilson}]{finzi-icml-2021}
Finzi, M., Welling, M., and Wilson, A.~G., \enquote{A practical method for
  constructing equivariant multilayer perceptrons for arbitrary matrix groups,}
  \emph{International conference on machine learning}, PMLR, 2021, pp.
  3318--3328.
\newblock \doi{10.48550/arXiv.2104.09459}.

\bibitem[{Wu et~al.(2020)Wu, Pan, Chen, Long, Zhang, and Philip}]{wu-ieee-2020}
Wu, Z., Pan, S., Chen, F., Long, G., Zhang, C., and Philip, S.~Y., \enquote{A
  comprehensive survey on graph neural networks,} \emph{IEEE transactions on
  neural networks and learning systems}, Vol.~32, No.~1, 2020, pp. 4--24.
\newblock \doi{10.1109/TNNLS.2020.2978386}.

\bibitem[{Keriven and Peyr{\'e}(2019)}]{keriven-anips-2019}
Keriven, N., and Peyr{\'e}, G., \enquote{Universal invariant and equivariant
  graph neural networks,} \emph{Advances in Neural Information Processing
  Systems}, Vol.~32, 2019.
\newblock \doi{10.48550/arXiv.1905.04943}.

\bibitem[{Succi(2018)}]{succi-book-2018}
Succi, S., \emph{{The Lattice {B}oltzmann Equation: For Complex States of
  Flowing Matter}}, OUP Oxford, 2018.
\newblock \doi{10.1093/oso/9780199592357.001.0001}.

\bibitem[{Hennigh(2017)}]{hennigh-arxiv-2017}
Hennigh, O., \enquote{Lat-net: compressing lattice Boltzmann flow simulations
  using deep neural networks,} \emph{arXiv preprint arXiv:1705.09036}, 2017.
\newblock \doi{10.48550/arXiv.1705.09036}.

\bibitem[{Guo et~al.(2016)Guo, Li, and Iorio}]{guo-proc-2016}
Guo, X., Li, W., and Iorio, F., \enquote{Convolutional Neural Networks for
  Steady Flow Approximation,} \emph{Proceedings of the 22nd ACM SIGKDD
  International Conference on Knowledge Discovery and Data Mining}, Association
  for Computing Machinery, New York, NY, USA, 2016, pp. 481--490.
\newblock \doi{10.1145/2939672.2939738}.

\bibitem[{Wang et~al.(2021)Wang, Chung, Armstrong, and
  Mostaghimi}]{wang-tpm-2021}
Wang, Y.~D., Chung, T., Armstrong, R.~T., and Mostaghimi, P., \enquote{ML-LBM:
  predicting and accelerating steady state flow simulation in porous media with
  convolutional neural networks,} \emph{Transport in Porous Media}, Vol. 138,
  No.~1, 2021, pp. 49--75.
\newblock \doi{10.1007/s11242-021-01590-6}.

\bibitem[{Corbetta et~al.(2023)Corbetta, Gabbana, Gyrya, Livescu, Prins, and
  Toschi}]{corbetta-epje-2023}
Corbetta, A., Gabbana, A., Gyrya, V., Livescu, D., Prins, J., and Toschi, F.,
  \enquote{Toward learning Lattice Boltzmann collision operators,} \emph{The
  European Physical Journal E}, Vol.~46, No.~3, 2023, p.~10.
\newblock \doi{10.1140/epje/s10189-023-00267-w}.

\bibitem[{Bedrunka et~al.(2021)Bedrunka, Wilde, Kliemank, Reith, Foysi, and
  Kr{\"a}mer}]{bedrunka-hpc-2021}
Bedrunka, M.~C., Wilde, D., Kliemank, M., Reith, D., Foysi, H., and Kr{\"a}mer,
  A., \enquote{Lettuce: PyTorch-Based Lattice Boltzmann Framework,} Springer
  International Publishing, Cham, 2021, pp. 40--55.
\newblock \doi{10.1007/978-3-030-90539-2_3}.

\bibitem[{Horstmann et~al.(2024)Horstmann, Bedrunka, and
  Foysi}]{horstmann-cf-2024}
Horstmann, J.~T., Bedrunka, M.~C., and Foysi, H., \enquote{Lattice Boltzmann
  method with artificial bulk viscosity using a neural collision operator,}
  \emph{Computers \& Fluids}, Vol. 272, 2024, p. 106191.
\newblock \doi{10.1016/j.compfluid.2024.106191}.

\bibitem[{Ortali et~al.(2024)Ortali, Gabbana, Demo, Rozza, and
  Toschi}]{ortali-arxiv-2024}
Ortali, G., Gabbana, A., Demo, N., Rozza, G., and Toschi, F., \enquote{Kinetic
  data-driven approach to turbulence subgrid modeling,} \emph{arXiv preprint
  arXiv:2403.18466}, 2024.
\newblock \doi{10.48550/arXiv.2403.18466}.

\bibitem[{Goodfellow et~al.(2016)Goodfellow, Bengio, and
  Courville}]{goodfellow-book-2016}
Goodfellow, I., Bengio, Y., and Courville, A., \emph{Deep Learning}, MIT Press,
  2016.
\newblock \urlprefix\url{http://www.deeplearningbook.org}.

\bibitem[{Kr\"uger et~al.(2017)Kr\"uger, Kusumaatmaja, Kuzmin, Shardt, Silva,
  and Viggen}]{kruger-book-2017}
Kr\"uger, T., Kusumaatmaja, H., Kuzmin, A., Shardt, O., Silva, G., and Viggen,
  E.~M., \emph{{The Lattice {B}oltzmann Method}}, Springer International
  Publishing, 2017.
\newblock \doi{10.1007/978-3-319-44649-3}.

\bibitem[{Shan(2016)}]{shan-jocs-2016}
Shan, X., \enquote{{The mathematical structure of the lattices of the lattice
  Boltzmann method},} \emph{Journal of Computational Science}, Vol.~17, 2016,
  pp. 475--481.
\newblock \doi{10.1016/j.jocs.2016.03.002}.

\bibitem[{Bhatnagar et~al.(1954)Bhatnagar, Gross, and
  Krook}]{bhatnagar-pr-1954}
Bhatnagar, P.~L., Gross, E.~P., and Krook, M., \enquote{{A Model for Collision
  Processes in Gases. Amplitude Processes in Charged and Neutral One-Component
  Systems},} \emph{Phys. Rev.}, Vol.~94, No.~3, 1954, pp. 511--525.
\newblock \doi{10.1103/PhysRev.94.511}.

\bibitem[{Artin(1998)}]{artin-book}
Artin, M., \emph{Algebra}, Birkh{\"a}user, 1998.

\bibitem[{LeCun et~al.(2015)LeCun, Bengio, and Hinton}]{lecun2015deep}
LeCun, Y., Bengio, Y., and Hinton, G., \enquote{Deep learning,} \emph{Nature},
  Vol. 521, No. 7553, 2015, pp. 436--444.
\newblock \doi{10.1038/nature14539}.

\bibitem[{Kingma and Ba(2014)}]{kingma-arxiv-2014}
Kingma, D.~P., and Ba, J., \enquote{Adam: A method for stochastic
  optimization,} \emph{arXiv preprint arXiv:1412.6980}, 2014.
\newblock \doi{10.48550/arXiv.1412.6980}.

\bibitem[{Bedolla et~al.(2020)Bedolla, Padierna, and
  Castaneda-Priego}]{bedolla-jop-2020}
Bedolla, E., Padierna, L.~C., and Castaneda-Priego, R., \enquote{Machine
  learning for condensed matter physics,} \emph{Journal of Physics: Condensed
  Matter}, Vol.~33, No.~5, 2020, p. 053001.
\newblock \doi{10.1088/1361-648X/abb895}.

\bibitem[{Keith et~al.(2021)Keith, Vassilev-Galindo, Cheng, Chmiela, Gastegger,
  Muller, and Tkatchenko}]{keith-cr-2021}
Keith, J.~A., Vassilev-Galindo, V., Cheng, B., Chmiela, S., Gastegger, M.,
  Muller, K.-R., and Tkatchenko, A., \enquote{Combining machine learning and
  computational chemistry for predictive insights into chemical systems,}
  \emph{Chemical reviews}, Vol. 121, No.~16, 2021, pp. 9816--9872.
\newblock \doi{10.1021/acs.chemrev.1c00107}.

\bibitem[{Boyda et~al.(2022)Boyda, Cal{\`\i}, Foreman, Funcke, Hackett, Lin,
  Aarts, Alexandru, Jin, Lucini et~al.}]{boyda-arxiv-2022}
Boyda, D., Cal{\`\i}, S., Foreman, S., Funcke, L., Hackett, D.~C., Lin, Y.,
  Aarts, G., Alexandru, A., Jin, X.-Y., Lucini, B., et~al.,
  \enquote{Applications of machine learning to lattice quantum field theory,}
  \emph{arXiv preprint arXiv:2202.05838}, 2022.
\newblock \doi{10.48550/arXiv.2202.05838}.

\bibitem[{Gabbana et~al.(2020)Gabbana, Simeoni, Succi, and
  Tripiccione}]{gabbana-pr-2020}
Gabbana, A., Simeoni, D., Succi, S., and Tripiccione, R., \enquote{Relativistic
  lattice Boltzmann methods: Theory and applications,} \emph{Physics Reports},
  Vol. 863, 2020, pp. 1 -- 63.
\newblock \doi{10.1016/j.physrep.2020.03.004}, relativistic lattice Boltzmann
  methods: Theory and applications.

\end{thebibliography}

\end{document}